\documentclass[12pt]{article}
\pdfoutput=1
\usepackage{graphicx} 
\usepackage[utf8]{inputenc}
\usepackage[margin=1in]{geometry} 
\usepackage{amsmath,amsthm,amssymb,hyperref}
\hypersetup{
    colorlinks=true,
    linkcolor=blue,
    citecolor=magenta,
    filecolor=magenta,      
    urlcolor=blue,
    pdftitle={},
    pdfpagemode=FullScreen,
    }
\usepackage{adjustbox}
\usepackage{caption}
\usepackage{float}
\usepackage{subcaption}
\usepackage{xcolor}
\usepackage{soul}
\usepackage{bbm}
\usepackage{enumitem}
\usepackage{physics}
\usepackage{amssymb}
\usepackage{tikz}

\usetikzlibrary{automata, positioning, arrows, calc}

\tikzset{
	<->,  
	>=stealth, 
	node distance=5cm, 
	every state/.style={draw=blue!55,very thick,fill=blue!20}, 
    initial/.style ={}
}

\usepackage[
    backend=biber,
    style=phys,
  ]{biblatex}
\numberwithin{theorem}{subsection}
\numberwithin{definition}{subsection}
\numberwithin{lemma}{subsection}

\newcommand{\beq}{\begin{equation}\begin{aligned}}
\newcommand{\eeq}{\end{aligned}\end{equation}}
\usepackage{stackengine}

\usepackage{authblk}
\usepackage{amsthm}

\theoremstyle{plain}

\theoremstyle{definition}

\newcommand{\Z}{{\mathbb Z}}

\title{Non-invertible SPT, gauging and symmetry fractionalization}

\author[1,2]{Yabo Li}
\author[1,2]{Mikhail Litvinov}
\affil[1]{\em{C. N. Yang Institute for Theoretical Physics, Stony Brook University}}
\affil[2]{\em{Department of Physics and Astronomy, Stony Brook University}}
\begin{document}

\begin{titlepage}
\maketitle

\begin{abstract}

    We explicitly realize the Rep($Q_8$) non-invertible symmetry-protected topological (SPT) state as a 1+1d cluster state on a tensor product Hilbert space of qubits.
    Using the Kramers-Wannier operator, we construct the lattice models for the phases of all the symmetries in the Rep($Q_8$) duality web.
    We further show that we can construct a class of lattice models with Rep($G$) symmetry including non-invertible SPT phases if they have a dual anomalous abelian symmetry.
    Upon dualizing, there is a rich interplay between onsite symmetries, non-onsite symmetries, non-abelian symmetries, and non-invertible symmetries.
    We show that these interplay can be explained using the symmetry fractionalization in the 2+1d bulk SET.
\end{abstract}
\end{titlepage}

\tableofcontents

\section{Introduction}
Symmetry-protected topological (SPT) states have been studied for over a decade~\cite{chen2011classification,turner2011topological,schuch2011classifying,lu2012theory,pollman2012aymmetry,chen2013symmetry,chen2014symmetry,senthil2015symmetry}, offering fascinating insights into quantum many-body physics.
These states, although requiring symmetries for their realization, can still be observed in nature.
They have found practical applications, notably in measurement-based quantum computation (MBQC), where they provide robust frameworks for quantum operations~\cite{raussendorf2002one,briegel2009measurement,miyake2010quantum,else2012symmetry,stephen2017computational,raussendorf2017symmetry,wei2017universal,wei2018quantum,raussendorf2019computationally}.

In recent years, there has been a surge of interest in non-invertible symmetries~\cite{mcgreevy2023generalized,cordova2022snowmass,brennan2023introduction,schafer2024ictp,shao2023s}, including the SPT phases under non-invertible symmetries~\cite{thorngren2019fusion,inamura2021topological,bhardwaj2024lattice}.
A slew of examples appeared recently in the literature realizing non-invertible symmetries on the tensor product Hilbert space~\cite{fechisin2023non,seifnashri2024cluster, bhardwaj2024illustrating}.
However, a comprehensive framework for writing explicit ultraviolet (UV) lattice realizations of these states on the tensor product Hilbert space remains largely undeveloped.

In this work, we continue studying Rep($G$) SPT states, or fiber functors as they are known in mathematics~\cite{tambara2000representations}.
We start with the duality frame for a Rep$(G)$ category, where the symmetry $G'$ is abelian potentially with some anomaly $H^3(G', U(1))$.
The abelian anomalous theories can be systematically constructed as a boundary of the 2+1d SPT on a tensor product Hilbert space.
We then gauge the abelian symmetries to obtain lattice models for phases under Rep($G$) symmetry or other Morita equivalent fusion categories.
If we start with certain spontaneous symmetry-breaking (SSB) phase permitted by the model's anomalies, we can obtain the non-invertible SPT phases after gauging.




We structured the paper as follows:
In Sec.~\ref{sec:repG}, we explore the explicit realization of Rep($G$) symmetries from gauging anomalous abelian symmetries.
In Sec.~\ref{sec:SymFrac}, the symmetry fractionalization in 2+1D SET phases is discussed, explaining how gauging subgroups of anomalous symmetries leads to emergent non-abelian or non-invertible symmetries on the boundary. 
In App.~\ref{app:PEM}, we explain how partial electric-magnetic duality relates different twisted quantum doubles.
In App.~\ref{app:Z_2^3}, it is shown how to obtain type II anomalies from the type III anomaly using the Levin-Gu model and the spontaneous symmetry breaking.
In App.~\ref{app:tachikawa}, we reproduce the relation between Tambara-Yamgami categories and $\mathrm{Vec}_{D_8}^\omega$ symmetries. 
We also write explicit forms of cocycles for the anomaly of a dual group of some Tambara-Yamagami categories.

\begin{figure}[h]
    \centering
    \includegraphics[width=0.7\linewidth]{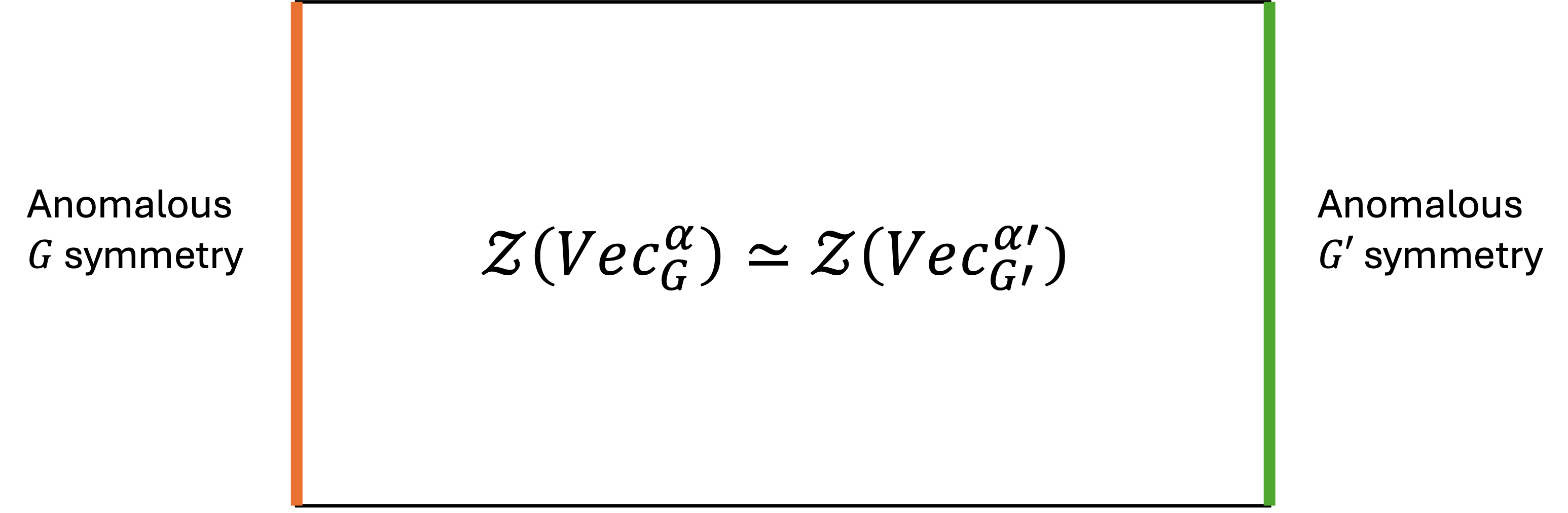}
    \caption{Some $G$ twisted quantum double and $G'$ twisted quantum double can be identical according to the PEM duality (more details in Sec.~\ref{app:PEM}).
    When $\alpha=1$, different gapped boundaries of $G$ quantum double and $G'$ twisted quantum double are all related to each other by gauging.
    For example, the non-invertible Rep($G$) symmetry is dual to an anomalous $G'$ symmetry as a result.}
    \label{fig:sym_tft}
\end{figure}

\section{Rep($G$) SPT phases}
\label{sec:repG}
It is known that different gapped boundaries of a topological order can be related to each other by gauging~\cite{kitaev2012models}.
For example, in 1+1d, $G$ symmetric phases and Rep($G$) symmetric phases are the smooth and rough boundaries of a $G$ quantum double, and they are dual to each other via the gauging of $G$ (or Rep($G$)).
In 2+1d, the partial electric-magnetic (PEM) duality shows that some twisted quantum double models in 2+1d are the same topological orders (see Appendix \ref{app:PEM}).

In this section, we show how different abelian symmetries can be dualized to non-abelian and non-invertible symmetries under the guidance of PEM duality (See Fig.~\ref{fig:sym_tft}). Starting from the anomalous symmetry operators obtained from bulk SPT, we construct the lattice models for all the dual symmetries, the most notable lattice models of which are the non-invertible Rep($G$) SPT phases.

\subsection{Rep($D_8$) symmetry}
According to the PEM duality, the 2+1d type III twisted quantum double of group $\hat{\Z}_2^e\times \hat{\Z}_2^o\times \hat{\Z}_2^v$, given by 3-cocycle\footnote{Definitions of type I, II, III cocycle of $\Z_2^3$ group is given in Appendix \ref{app:Z_2^3}.}
\beq
    \alpha=\frac{1}{2}A_e\cup A_o\cup A_v,
\eeq
are PEM dual to the quantum double of an extension group of $\hat{\Z}_2^e\times \hat{\Z}_2^o$ by $\Z_2$. The group extension is given by a 2-cocycle $[\hat{F}]\in H^2(\hat{\Z}_2^e\times \hat{\Z}_2^o,\Z_2)$,
\beq
    \hat{F}=\frac{1}{2}A_e\cup A_o.
\eeq
This group is $D_8$.
We can also gauge the full category to obtain $Rep(D_8)$.
From symTFT, we conclude that the 1+1d symmetry $\mathrm{Vec}_{\Z_2^3}^{III}$ is dual to both $\mathrm{Vec}_{D_8}$ and Rep($D_8$).
Details are left for the Appendix \ref{app:PEM}.

\subsubsection{Rep($D_8$) SPT models}
We start by presenting the explicit realization of the Rep($D_8$) category on the lattice from Ref.~\cite{seifnashri2024cluster}.
The three Rep($D_8$) SPT models can be given from twisted gauging $\hat{\Z}_2^e\times \hat{\Z}_2^o$ from different symmetry breaking phases under type III anomalous $\Z_2^3$ symmetry.

In this paper, we always choose the lattice realization to be a chain with $L = 4N $ sites, and use the periodic boundary condition if not specified otherwise. For Rep($D_8$) symmetry there are three symmetries operators\footnote{Note that this realization is consistent with the dual frame presentation of the $\Z_2^3$ boundary model of the Levin-Gu SPT.}: $\eta_e = \prod_j X_{2j}$, $\eta_o = \prod_j X_{2j+1}$ and 
$\mathsf{D}=T \mathsf{D}_e \mathsf{D}_o$,
where $T$ is the lattice translation by one site, and $\mathsf{D}_e$ and $\mathsf{D}_o$ are the Kramers-Wannier operators on even and odd sites~\cite{seifnashri2024cluster},
\begin{equation}
\begin{aligned}
& \mathsf{D}_e=e^{\frac{2 \pi i L}{16}} \frac{1+\eta_e}{\sqrt{2}} \frac{1-i X_L}{\sqrt{2}} \cdots \frac{1-i Z_4 Z_2}{\sqrt{2}} \frac{1-i X_2}{\sqrt{2}}, \\
& \mathsf{D}_o=e^{\frac{2 \pi i L}{16}} \frac{1+\eta_o}{\sqrt{2}} \frac{1-i X_{L-1}}{\sqrt{2}} \cdots \frac{1-i Z_3 Z_1}{\sqrt{2}} \frac{1-i X_1}{\sqrt{2}},
\end{aligned}
\end{equation}
the square of which are given by $\mathsf{D}_e^2 = T_e^{-1}(1+\eta_e)$, $\mathsf{D}_o^2 = T_o^{-1}(1+\eta_o)$. The $\mathsf{D}$ operator acts on Pauli operators as
\begin{equation}
    X_j \rightsquigarrow Z_{j-1} Z_{j+1}, \quad Z_{j-1} Z_{j+1} \rightsquigarrow X_j.
\end{equation}
The operators $\eta_e, \eta_o$, and $\mathsf{D}$ satisfy the following Rep($D_8$) fusion rules
\begin{equation}
\begin{aligned}
& \mathsf{D}^2=1+\eta_e+\eta_o+\eta_e \eta_o \\
& \eta_e \mathsf{D}=\mathsf{D} \eta_e=\eta_o \mathsf{D}=\mathsf{D} \eta_o=\mathsf{D}.
\end{aligned}
\end{equation}

\subsubsection{Form Rep($D_8$) to $D_8$ in one shot}
We can directly obtain a $D_8$ symmetry by gauging a $\Z_2$ subgroup of the Rep($D_8$) category (see Appendix \ref{app:tachikawa}). We choose to gauge the $\eta_e$ symmetry in this section, with the help of the Kramers-Wannier operator $\mathsf{D}_e$. After gauging, one of the symmetries in this phase can be obtained by considering a Wilson line of a dual symmetry: $    \check{\eta}_e=\prod_{j} \check{X}_{2j}$. The other two symmetry operators can be obtained from pushing $\eta_o$ and $\mathsf{D}$ through $\mathsf{D}_e$. Since $\eta_o$ commutes with $\mathsf{D}$, after gauging $\Z_2^e$,
\beq
    \check{\eta}_o=\prod_j \check{X}_{2j+1}.
\eeq 
We can find the last symmetry operator by pushing the $\mathsf{D}$ operator through $\mathsf{D}$ as follows: 
\begin{equation}
    \mathsf{D}_e T \mathsf{D}_e \mathsf{D}_o = T \mathsf{D}_o \mathsf{D}_e \mathsf{D}_o =T \mathsf{D}_o^2 \mathsf{D}_e = T T_o^{-1}(1 + \eta_o) \mathsf{D}_e.
    \label{eq:SwapD8}
\end{equation}
Let us consider any term in the Hamiltonian invariant under Rep($D_8$): $\mathcal{O}(X_j,Z_j Z_{j+2})$.
By employing the Eq.~\eqref{eq:SwapD8}, we see that operator\footnote{$B $ doesn't have a projector in the definition as this relation is true on the symmetric subspace. }
\begin{equation}
   B = T T_o ^{-1} = \prod_j \text{Swap}_{2j,2j+1}
\end{equation}
commutes with the $\mathsf{D}_e\mathcal{O}\mathsf{D}_e$ operator,
\beq
    B \mathsf{D}_e \mathcal{O} \mathsf{D}_e =\frac{1}{2} \mathsf{D}_e \mathsf{D} \mathcal{O} \mathsf{D}_e = \frac{1}{2} \mathsf{D}_e \mathcal{O} \mathsf{D} \mathsf{D}_e = \mathsf{D}_e \mathcal{O} \mathsf{D}_e B.
\eeq
Therefore, $B$ will be a symmetry of any gauged Hamiltonian.

The group $D_8$ is generated by two elements $\langle a,x\rangle$ satisfying the following relations,
$xax=a^3$, $x^2=1$, $a^4=1$. Any element $\tilde{g}\in D_8$ can be written as
\beq
    \tilde{g}=x^G a^g=:(G,g),
    \label{eq:D_8}
\eeq
where $G=0,1$ and $g=0,1,2,3$. The group multiplication is given by
\beq
(G,g)\cdot(H,h)=x^G a^g \cdot x^H a^h=x^{G+H}a^{(-1)^H g+h}=([G+H]_2,[(-1)^H g+h]_4).
\eeq

Under the convention in Eq.~\eqref{eq:D_8}, $\check{\eta}_o$, $\check{\eta}_e$ and $B$ form a $D_8$ symmetry group with the following embedding:
\begin{equation}
    \check{\eta}_e = xa, \quad \check{\eta}_o = xa^3, \quad B = x.
\end{equation}

Therefore, we conclude the duality web for Rep($D_8$) symmetry as follows\footnote{
We note that this duality web was also discussed in the continuum in Ref.~\cite{bhardwaj2018finite,diatlyk2024gauging}.}, 

\begin{equation}
    \begin{tikzpicture}
		\node[initial] (s0) {Rep($D_8$)};
		\node[right of=s0] (s1) {$\mathrm{Vec}_{D_8}$};
		\node[right of=s1] (s2) {$\mathrm{Vec}_{\hat{\Z}_2^e\times \hat{\Z}_2^o\times \hat{\Z}_2^v}^{III}$};
		\node[right of=s2] (s3) {$\mathrm{Vec}_{D_8}$};
		
		\draw           
                (s0) edge[->] node[above]{Gauge $\Z_2^e$} (s1)
                (s1) edge[<-] node[above]{Gauge $\hat{\Z}_2^o$} (s2)
                (s2) edge[->] node[above]{Gauge $\hat{\Z}_2^v$} (s3);
	\end{tikzpicture}
\end{equation}

What happens to the three Rep($D_8$) SPT states after gauging?
The cluster state turns into the following Hamiltonian 
\begin{equation}
    \check{H} = -\sum_{j=1}^{L/2} (\check{Z}_{2j}\check{Z}_{2j+1}\check{Z}_{2j+2}\check{Z}_{2j+3}+ \check{X}_{2j}\check{X}_{2j+1}).
\end{equation}
The ground states of this Hamiltonian are given by
\begin{equation}
    \bigotimes_{j=1}^{L/2} \frac{1}{\sqrt{2}}\Big(\ket{00}+\ket{11}\Big)_{2j,2j+1}, \quad \bigotimes_{j=1}^{L/2}\frac{1}{\sqrt{2}}\Big(\ket{01}+ \ket{10}\Big)_{2j,2j+1},
\end{equation}
where we see that both $\check{\eta}_o$ and $\check{\eta}_e$ are broken and the diagonal subgroup $\check{\eta}_o \check{\eta}_e $ is preserved.
The order parameter of the phase is $\check{Z}_{2j} \check{Z}_{2j+1}$, which is invariant under $\check{\eta}_o \check{\eta}_e $ and the swap symmetry.
All possible phases with symmetries $G$ in 1+1d are classified by tuples $(H, A)$, where $H$ is the unbroken subgroup of $G$, and $A\in H^2(H,U(1))$\cite{schuch2011classifying}\footnote{There is no canonical choice of a trivial SPT state, and usually it's only meaningful to consider their difference. We choose a trivial product state to be the trivial SPT state and consider differences with respect to it.}.
So, this phase corresponds to the breaking pattern $D_8 \to ( \Z_2\times \Z_2, 0)$. 

If we take the Hamiltonian for the `odd' Rep($D_8$) SPT~\cite{seifnashri2024cluster},
\beq
    H_{\text {odd }}=\sum_{j=1}^{L / 2} Z_{2 j-1} X_{2 j} Z_{2 j+1}- Y_{2 j} X_{2 j+1} Y_{2 j+2} (1+ Z_{2 j-1} X_{2 j} X_{2 j+2} Z_{2 j+3}),
\eeq
after gauging $\eta_e$ we find the following Hamiltonian for the $D_8$ phase:
\begin{equation}
    \check{H}_{\text {odd }}=\sum_{j=1}^{L/2} \check{Z}_{2j} \check{Z}_{2j+1} \check{Z}_{2j+2}\check{Z}_{2j+3} + \check{Z}_{2j}\check{X}_{2j+2} \check{X}_{2j+3}\check{Z}_{2j+4}+ \check{Z}_{2j+1}\check{X}_{2j+2}\check{X}_{2j+3}\check{Z}_{2j+5}.
\end{equation}
Notice that the last two terms are interchanged by the swap symmetry. It can be showed that this is in a different symmetry-protected phase with respect to the unbroken $(\Z_2 \times \Z_2)$ group: $(\Z_2\times\Z_2, 1)$.

A similar-looking Hamiltonian is obtained for the `even' case:
\begin{equation}
    \check{H}_{\mathrm{even}}=\sum_{j=1}^{L/2} \check{X}_{2j} \check{X}_{2j+1}- \check{Z}_{2j}\check{Y}_{2j+1} \check{Z}_{2j+2}\check{Y}_{2j+3}(1+\check{X}_{2j}\check{X}_{2j+2}).
\end{equation}
The symmetry is generated by $\langle1, B\check{\eta}_o,\check{\eta}_e\check{\eta}_o, B\check{\eta}_e\rangle$ and forms group $\Z_4$.

As a side note, we can consider a non-trivial SPT state for the $D_8$ symmetry.
To find it explicitly, one might consider a slightly corrected cluster state Hamiltonian to account for the swap symmetry.
One of the representatives is
\begin{equation}
    H = -\sum_{j=1}^{L/2} (Z_{2j-1}X_{2j}Z_{2j+3} + Z_{2j-2} X_{2j+1}Z_{2j+2}).
\end{equation}
It is trivially invariant and is also an SPT for $\Z_2\times \Z_2$ subgroup $\eta_0,\eta_e$.

\subsection{Rep($Q_8$) symmetry}
In this section, we discuss the whole duality web of Rep($Q_8$) symmetry, including Rep($Q_8$), $Q_8$, type II+II+III anomalous $\Z_2^3$, and anomalous $D_8$. We give the lattice Hamiltonians for the fixed-point states of all the phases in the duality web.
\subsubsection{$\Z_2^3$ type II+II+III anomaly}
From the discussion in Appendix \ref{app:PEM}, the type II+II+III twisted quantum double given by 3-cocycle 
\beq
    \frac{1}{2} A_e\cup A_e \cup A_v+\frac{1}{2} A_o\cup A_o\cup A_v+\frac{1}{2} A_e\cup A_o\cup A_v,
    \label{eq:Q8anom}
\eeq
is PEM dual to the quantum double of $Q_8$. Thus we know that there exists a duality between $\Z_2^3$ type II+II+III anomalous symmetry and Rep($Q_8$) symmetry. 
Furthermore, from the derivations in Appendix \ref{app:Z_2^3}, the type II+II+III anomalous symmetry operators of $\Z_2^3=\hat{\Z}_2^e\times \hat{\Z}_2^o\times \hat{\Z}_2^v$ on a 1d chain are given by
\beq
    \hat{U}_e=\prod_j \hat{X}_{2j},\ \hat{U}_o=\prod_{j}\hat{X}_{2j+1},\ \hat{V}=\prod_{j} \underbrace{C\hat{Z}_{j,j+1}}_{\text{type III}}\underbrace{\hat{Z}_j C\hat{Z}_{j,j+2}}_{\text{type II+II}},
    \label{eq:II+II+III ops}
\eeq
where $ C\hat{Z}_{j,j+1} $ in $\hat{V}$ is a part giving a type III anomaly appearing at the boundary of the Levin-Gu model~\cite{levin2012braiding,yoshida2017gapped}. The Kennedy-Tasaki operator,
\beq
    \mathsf{KT}=(\prod_j CZ_{j,j+1})\mathsf{D}(\prod_j CZ_{j,j+1}),
\eeq
has actions on the operators as follows,
\begin{equation}
    X_j \rightsquigarrow X_j,\quad
    Z_{j-1}Z_{j+1} \rightsquigarrow Z_{j-1}X_jZ_{j+1}.
\end{equation}

By applying the $\mathsf{KT}$ operator, we implement a twisted gauging. After such twisted gauging, we have two emergent dual symmetry operators
\beq
    U_e=\prod_j X_{2j},\ U_o=\prod_j X_{2j+1}.
\eeq
The other symmetry operator can be obtained by pushing the unitary $\hat{V}$ through the $\mathsf{KT}$, 
\beq
    & 2\mathsf{KT}\cdot \hat{V}=\mathsf{D}'\cdot \mathsf{KT}, \\
    & \mathsf{D}'\equiv e^{\frac{2\pi i L}{8}}(\prod_j \frac{1+iZ_{j-1}X_j Z_{j+1}}{\sqrt{2}})\mathsf{D},
\eeq
where we used that $(\prod_j \frac{1+iX_j}{\sqrt{2}}) Z_k = Y_k (\prod_j \frac{1+iX_j}{\sqrt{2}})$ and $(\prod_j \frac{1+iX_j}{\sqrt{2}}) Y_k = Z_k (\prod_j \frac{1+iX_j}{\sqrt{2}})$. The non-invertible operator $\mathsf{D'}$ acts on the operators in the following way:
\begin{equation}
    \begin{split}
        X_j \rightsquigarrow Z_{j-2}Y_{j-1} Y_{j+1} Z_{j+2},\quad       
        Z_{j-1}Z_{j+1}\rightsquigarrow -Z_{j-2}X_{j-1}X_{j}X_{j+1}Z_{j+2}.
    \end{split}
\end{equation}
The fusion rules of the above symmetry operators are given by
\beq
     \mathsf{D}'U_e=\mathsf{D}'U_o=U_e\mathsf{D}'=U_o\mathsf{D}'=\mathsf{D}',\ \mathsf{D}'^2=1+U_e+U_o+U_e U_o.
\eeq
Therefore we show on lattice that, after twisted gauging the $\hat{\Z}_2^e\times \hat{\Z}_2^o$ symmetry from type II+II+III anomaly, we do obtain a Rep($Q_8$) symmetry\footnote{To be more precise, the fusion rules are not enough to tell Rep($Q_8$) and Rep($D_8$) apart, and we need to consider the full fusion category (see \cite{seifnashri2023lieb,kawagoe2021anomalies} for procedures). We can avoid this problem by gauging $\hat{\Z}_2^v$ and showing that the dual symmetry is $Q_8$.}. 
Notice that the symmetry is fully internal and doesn't mix with the lattice transition. 
So, the symmetry category is realized exactly contrary to the critical Ising model \cite{seiberg2024majorana}.

\subsubsection{Rep($Q_8$) SPT model}
In this section, we construct the Rep($Q_8$) SPT state from twisted gauging a $\Z_2^3$ symmetry breaking phase.

To obtain the Rep($Q_8$) SPT state, we need to consider the symmetry breaking phase of $\hat{\Z}_2^e\times \hat{\Z}_2^o$, such that after gauging, the state is symmetric under the dual $\Z_2^e\times \Z_2^o$ symmetry. For anomalous type II+II+III symmetry in Eq.~\eqref{eq:Q8anom}, there is a mixed anomaly between $\hat{\Z}_2^v$ and $\hat{\Z}_2^e$, and also between $\hat{\Z}_2^v$ and $\hat{\Z}_2^o$. Because of the fact that when two symmetries are mixed anomalous, the ground states are in general constrained to be either in gapless phases or the symmetry breaking phases, the only possible gapped phase is the $\Z_2^e\times \Z_2^o$ broken and $\hat{\Z}_2^v$ unbroken phase:

\begin{equation}
    \hat{H} = -\sum_{j=1}^{L} \hat{Z}_{j-1}\hat{Z}_{j+1}.
\end{equation}

After twisted gauging $\hat{\Z}_2^e\times \hat{\Z}_2^o$, we obtain a cluster Hamiltonian,
\beq
    H=-\sum_{j=1}^{L} Z_{j-1} X_j Z_{j+1}.
    \label{eq:RepQ_8 SPT}
\eeq

Therefore, the cluster state is a Rep($Q_8$) SPT, with symmetry operators being $U_e$, $U_o$ and $\mathsf{KT}'$. We note that we can always apply on the SPT state a unitary gate that is symmetric under $\Z_2^e\times \Z_2^o$, such that the non-invertible operator is conjugated. For example, when we apply a cluster entangler $\prod_j CZ_{j,j+1}$, the SPT state is mapped from cluster state to trivial product state, while the non-invertible operator is conjugated to $\mathsf{D}''\equiv e^{\frac{2\pi i L}{8}}(\prod_i \frac{1+iX_i}{\sqrt{2}})\mathsf{KT}$. As a result, we can equivalently say that the trivial product state is a Rep($Q_8$) SPT state, with symmetry operators being $U_e$, $U_o$, and $\mathsf{D}''$. The conjugation of cluster entangler is essentially stacking with an $\Z_2\times \Z_2$ SPT on top of the Rep($Q_8$) SPT in Eq.~\eqref{eq:RepQ_8 SPT}. Indeed, if we chose to gauge $\hat{\Z}_2^e\times \hat{\Z}_2^o$ from the type II+II+III SSB phase, instead of twisted gauging, we would obtain a trivial state and non-invertible operator $\mathsf{KT}''$. Since the result of gauging and twisted gauging just differ by a stacking of normal SPT, we will use gauging to obtain non-invertible SPT states in the following sections for simplicity.

\subsubsection{Gauging Rep($Q_8$) symmetry}
In this section, we show that the Rep($Q_8$) symmetry and anomalous type II+II+III $\Z_2^3$ symmetry are dual to $Q_8$ symmetry and anomalous $D_8$ symmetry via gauging.

We start from considering the $\hat{\Z}_2^v$ symmetry operator in the type II+II+III anomalous $\Z_2^3$,
\beq
\hat{V}=\prod_{j}CZ_{j,j+1}Z_j CZ_{j,j+2},
\eeq
which is non-onsite\footnote{We drop the hats for all the type II+II+III symmetry operators in this section for simplicity.}. To gauge this symmetry, we enlarge our Hilbert space by introducing an auxiliary qubit on each site and then enforce gauge constraints~\cite{seifnashri2023lieb}.
We denote extra qubits with tildes.
Gauss's law (gauge constraint) takes the following form\footnote{The symmetry operator $\hat{V}$ is composed of commuting local gates, thus we can treat it as a width one symmetry.}:
\begin{equation}
    \tilde{X}_{j-1} CZ_{j-1, j} CZ_{{j-2},j} Z_{j} \tilde{X}_{j} .
\end{equation}
One of the symmetry operators after gauging $\hat{\Z}_2^v$ is the Wilson line of the dual $\Z_2$ group. The other symmetries of the model after gauging can be recovered by ``minimal coupling'' symmetry operators from the ungauged model to commute with Gauss's law. In summary, we can find a dual $Q_8$ symmetry is generated by the operators 
\beq
    \tilde{\eta}_e &= \prod_{j}X_{2j}CZ_{2j-1,\tilde{2j-1}}CZ_{2j,\tilde{2j}}CZ_{2j,\tilde{2j+1}},
    \\
    \tilde{\eta}_o &= \prod_{j} X_{2j+1}CZ_{2j-1,\tilde{2j-1}}CZ_{2j,\tilde{2j}}CZ_{2j+1,\tilde{2j+2}},
    \\
    \tilde{V} &= \prod_j \tilde{Z}_j,
\eeq
where we define $CZ_{i,\tilde{j}}:=\frac{1+Z_i+\tilde{Z}_j-Z_i\tilde{Z}_j}{2}$.
It is straightforward to check that they commute with the Gauss's law and generate a $Q_8$ symmetry since $\tilde{\eta}_e^2=\tilde{\eta}_o^2=\tilde{V}$, $\tilde{V}^2=1$ and $\tilde{\eta}_e \tilde{\eta}_o=\tilde{\eta}_o \tilde{\eta}_e \tilde{V}$.

Now suppose, instead of gauging $\hat{\Z}_2^v$, we gauge $\hat{\Z}_2^o$ from type II+II+III $\Z_2^3$. From the anomalous symmetry operators in Eq.~\eqref{eq:II+II+III ops}, we can obtain the dual symmetry after gauging using the Kramers-Wannier operator $\mathsf{D}_o$, 
\beq
    \check{\eta}_e&=\prod_j X_{2j},\quad
    \check{\eta}_o=\prod_j X_{2j+1},\\
    \check{V} &=\prod_j e^{\frac{2\pi i L}{8}} Z_{2j}CZ_{2j,2j+2}CX_{2j,2j+1}\Big(\frac{1+iX_{2j+1}}{\sqrt{2}}\Big),
    \label{eq:D_8 from Q_8}
\eeq
where $\check{\eta}_o$ is the dual symmetry operator, $\check{\eta}_e$ and $\check{V}$ are the results of pushing $\hat{\eta}_e$ and $\hat{V}$ through $\mathsf{D}_o$.
Under the convention in Eq.~\eqref{eq:D_8}, they form a $D_8$ symmetry group with the following embedding:
\begin{equation}
    \check{\eta}_e = x, \quad \check{\eta}_o = a^2, \quad \check{V} = a.
    \label{eq:D_8 anomalou embedding}
\end{equation}

Therefore, from gauging type II+II+III symmetry, we can obtain an $D_8$ symmetry. In the next section, we will show that the emergence of this anomalous $D_8$ can be understood from the symmetry fractionalization in the bulk SET. Besides, the bulk SET also has a $\Z_2^v\times \Z_2^e$ SPT in the bulk inherited from the type II+II+III SPT as in Eq.~\eqref{eq:bulk SET defectification}. If we choose the embedding as in Eq.~\eqref{eq:D_8 anomalou embedding}, and recall that the definition of $D_8$ group elements is in Eq.~\eqref{eq:D_8}, then the above cocycle gives a pull back 3-cocycle in $H^3(D_8,U(1))$ as
\beq
    \omega(\tilde{g},\tilde{h},\tilde{k})=(-1)^{gHK}.
    \label{eq:Q_8_cocycle}
\eeq

We conclude the duality web for Rep($Q_8$) symmetry as follows,

\begin{equation}
    \begin{tikzpicture}
		\node[initial] (s0) {Rep($Q_8$)};
		\node[right of=s0] (s1) {$\mathrm{Vec}_{D_8}^{\omega}$};
		\node[right of=s1] (s2) {$\mathrm{Vec}_{\hat{\Z}_2^e\times \hat{\Z}_2^o\times \hat{\Z}_2^v}^{II+II+III}$};
		\node[right of=s2] (s3) {$\mathrm{Vec}_{D_8}$};
		
		\draw           
                (s0) edge[<-] node[above]{Gauge $\check{\Z}_2^o$} (s1)
                (s1) edge[<-] node[above]{Gauge $\hat{\Z}_2^e$} (s2)
                (s2) edge[->] node[above]{Gauge $\hat{\Z}_2^v$} (s3);
	\end{tikzpicture}
\end{equation}

The fact that the Drinfeld centers $\mathcal{Z}(\mathrm{Rep}(Q_8))$ and $\mathcal{Z}(\mathrm{Vec}_{D_8}^{\omega})$ are identical was also obtained in Sec.~5.6 of Ref.~\cite{bhardwaj2018finite}. More details will be discussed in Appendix \ref{app:tachikawa}.

\subsection{Rep($G_1$) symmetry}
In this section, we discuss the SPT Hamiltonians of the representation category of the Pauli group. According to the PEM duality analysis in Appendix \ref{app:PEM}, the twisted quantum double of $\Z_2^4$, with a type II+III topological action
\beq
    \alpha=\frac{1}{2} A_1 \cup A_2 \cup A_4+ \frac{1}{2} A_3 \cup A_3\cup A_4,
\eeq
is equivalent to the quantum double of a group extension given by $[\hat{F}]\in H^2(\Z_2^3,\Z_2)$,
\beq
    \hat{F}=\frac{1}{2} A_1 \cup A_2+ \frac{1}{2} A_3 \cup A_3.
\eeq
This group is called a Pauli group, an order $16$ group formed by Pauli matrices,
\beq
    G_1=\{1,i,-1,-i,x,ix,-x,-ix,y,iy,-y,-iy,z,iz,-z,-iz\}.
\eeq
From the bulk SPT lattice model, we can obtain lattice realization, where on the qubit chain, each odd site hosts two qubit (type $o$ and type $o'$), each even site hosts one qubit (type $e$). The anomalous $\Z_2^4$ symmetry operators on this chain are
\beq
    \hat{U}_1=\prod_j \hat{X}_{2j},\ \hat{U}_2=\prod_{j} \hat{X}_{2j+1},\ \hat{U}_3=\prod_j \hat{X}'_{2j+1},\ \hat{V}=\prod_{j}\hat{Z}'_{2j-1}C'\hat{Z}'_{2j-1,2j+1}C\hat{Z}_{2j-1,2j+2},
\eeq
where $\hat{X}_{2j-1}'$ and $\hat{Z}_{2j-1}'$ are Pauli operators on the $o'$ qubit on site $2j-1$, and $C'\hat{Z}'_{2j-1,2j+1}:=\frac{1+\hat{Z}'_{2j-1}+\hat{Z}'_{2j+1}-\hat{Z}'_{2j-1}\hat{Z}'_{2j+2}}{2}$.

To obtain a Rep($G_1$) SPT, we should gauge the three onsite symmetries from a symmetry-breaking phase.
Now because $\hat{\Z}_2^{(4)}$ has mixed anomaly with both $\hat{\Z}_2^{(3)}$ and $\hat{\Z}_2^{(1)}\times \hat{\Z}_2^{(2)}$, there are only three possible symmetry breaking phases, corresponding to $\hat{\Z}_2^{(4)}$ unbroken, diag$(\hat{\Z}_2^{(1)}\times \hat{\Z}_2^{(4)})$ unbroken, and diag$(\hat{\Z}_2^{(1)}\times \hat{\Z}_2^{(4)})$ unbroken, respectively.
Therefore, after gauging $\hat{\Z}_2^{(1)}\times \hat{\Z}_2^{(2)}\times\hat{\Z}_2^{(3)}$ we expect three Rep($G_1$) SPT phases. The Rep($G_1$) symmetry operators are given by
\beq
    U_1&=\prod_j X_{2j},\ U_2=\prod_j X_{2j+1},\ U_3=\prod_j X'_{2j+1},\\
    \mathsf{D}&=e^{\frac{2\pi i L}{16}}\Big(\prod_j \frac{1+i X'_{2j+1}}{\sqrt{2}}\Big)\mathsf{D}_e \mathsf{D}_o,\\
    \mathsf{D}'&=U_3\mathsf{D}= e^{\frac{2\pi i L}{16}}\Big(\prod_j \frac{1-i X'_{2j+1}}{\sqrt{2}}\Big)\mathsf{D}_e \mathsf{D}_o.
\eeq
 The Rep($G_1$) fusion rules are given by
\beq
    U_1^2=U_2^2=U_3^2=1,\quad    U_3\mathsf{D}=\mathsf{D}U_3=\mathsf{D}',\quad U_1\mathsf{D}=\mathsf{D}U_1=U_2\mathsf{D}=\mathsf{D}U_2=\mathsf{D},\\
    \mathsf{D}^2=\mathsf{D}'^2=1+U_1+U_2+U_1 U_2,\quad U_1\mathsf{D}'=\mathsf{D}'U_1=U_2\mathsf{D}'=\mathsf{D}'U_2=\mathsf{D}.
\eeq
We can see from above that $U_1$, $U_2$ and $\mathsf{D}$ have the fusion rules of a Rep($D_8$) category. It turns out that the three Rep($G_1$) SPT states are related to Rep($D_8$) SPT states in Ref.~\cite{seifnashri2024cluster} as follows,
\beq
    \ket{\text{Rep}(G_1),\alpha}=\ket{\text{Rep}(D_8),\alpha}\otimes\ket{\text{product state}}_{o'}, \ \ \alpha=1,2,3.
\eeq

\subsection{Rep($G_{4,4}$) symmetry}
In this section, we discuss the $G_{4,4}$ and Rep($G_{4,4}$) symmetry from gauging $\Z_2^4$ and list all the non-invertible SPT lattice Hamiltonians.

We consider a $\Z_2^4=\hat{\Z}_2^o \times \hat{\Z}_2^{o'} \times \hat{\Z}_2^e\times \hat{\Z}_2^v$ anomaly given by,
\beq
    \frac{1}{2}A_o\cup A_e\cup A_v+\frac{1}{2}A_{o'}\cup A_e\cup A_v.
\eeq

Now if we gauge the $\hat{\Z}_2^v$ symmetry, from the PEM duality analysis in Appendix \ref{app:PEM}, we obtain the symmetry algebra of a non-abelian group $G_{4,4}=(\Z_2^2)\rtimes \Z_4$,
\beq
    \Tilde{\eta}_o^2=\Tilde{\eta}_{o'}^2=\Tilde{\eta}_e^2=\Tilde{V}^2=1,\ \Tilde{\eta}_o \Tilde{\eta}_e=\Tilde{V} \Tilde{\eta}_e \Tilde{\eta}_o,\ \Tilde{\eta}_{o'} \Tilde{\eta}_e=\Tilde{V} \Tilde{\eta}_e \Tilde{\eta}_{o'},\ \Tilde{\eta}_o \Tilde{\eta}_{o'}=\Tilde{\eta}_{o'} \Tilde{\eta}_o.
\eeq

Suppose we gauge instead the $\hat{\Z}_2^{o}\times \hat{\Z}_2^{o'}\times \hat{\Z}_2^{e}$ symmetry, we would have a non-invertible symmetry Rep($G_{4,4}$). For the lattice construction, on the qubit chain, each odd site hosts two qubits, and each even site hosts one qubit. The $\Z_2^4$ anomalous operators is given by,
\beq
    \hat{U}_o=\prod_j \hat{X}_{2j+1},\ \hat{U}_{o'}=\prod_j \hat{X}'_{2j+1},\ \hat{U}_e=\prod_j \hat{X}_{2j},\ \hat{V}=\prod_j C\hat{Z}_{2j,2j-1}C\hat{Z}_{2j,2j+1}C\hat{Z}'_{2j,2j-1}C\hat{Z}'_{2j,2j+1}.
    \label{eq:G44 sym op}
\eeq
After gauging the first three symmetries by $\mathsf{D}_e \mathsf{D}_o\mathsf{D}_{o'}$, the Rep($G_{4,4}$) symmetry are given by the dual onsite operators,
\beq
    U_o=\prod_{j} X_{2j+1},\ U_{o'}=\prod_{j}X'_{2j+1},\ U_e=\prod_{j}X_{2j},
\eeq
and a non-invertible operator 
\beq
    \mathsf{KT}'=(\mathsf{D}_e \mathsf{D}_o\mathsf{D}_{o'})^{\dagger}\hat{V}\mathsf{D}_e \mathsf{D}_o\mathsf{D}_{o'},
\eeq
that implements the following operator mapping,
\beq
    Z_{2j}Z_{2j+2}\rightsquigarrow Z_{2j-1}X_{2j}Z_{2j+1},\ Z'_{2j}Z'_{2j+2}\rightsquigarrow Z'_{2j-1}X_{2j}Z'_{2j+1},\\ 
    Z_{2j-1}Z_{2j+1}\rightsquigarrow Z_{2j-2}X_{2j-2}X'_{2j-1}Z_{2j},\ X_{j}\rightsquigarrow X_j,\ X'_{2j}\rightsquigarrow X'_{2j}.
\eeq

\subsubsection{Rep($G_{4,4}$) SPT models}

To obtain the SPT phases, we need to start from the phases where $\hat{\Z}_2^{o}$, $\hat{\Z}_2^{o'}$, and $\hat{\Z}_2^{e}$ are all broken. The possible symmetry breaking phases can be characterized by the unbroken $\Z_2$ group: 
\beq
    \hat{\Z}_2^{v},\ \mathrm{diag}(\hat{\Z}_2^{o}\times \hat{\Z}_2^{v}),\ \mathrm{diag}(\hat{\Z}_2^{o'}\times \hat{\Z}_2^{v}),\ \mathrm{diag}(\hat{\Z}_2^{e}\times \hat{\Z}_2^{v}),\\ \mathrm{diag}(\hat{\Z}_2^{o}\times \hat{\Z}_2^{o'}\times \hat{\Z}_2^{v}),\ \mathrm{diag}(\hat{\Z}_2^{o}\times \hat{\Z}_2^{o'}\times \hat{\Z}_2^{e}\times \hat{\Z}_2^{v}).
\eeq
Therefore, we know that there are 6 Rep($G_{4,4}$) SPT phases. It turns out that the SPT states fall into three types.

\paragraph{1. Rep($G_{4,4}$) SPT from $\Z_2^4/\hat{\Z}_2^{v}$ SSB.} For this phase, we start from the conventional GHZ Hamiltonian for the three onsite symmetries. After gauging $\hat{\Z}_2^{o}\times\hat{\Z}_2^{o'}\times\hat{\Z}_2^{e}$, we obtain a product state Hamiltonian 
\beq
    H_1=-\sum_{j=1}^{L/2}(X_{2j+1}+X'_{2j+1}+X_{2j}).
\eeq

\paragraph{2. Rep($G_{4,4}$) SPT from $\Z_2^4/\mathrm{diag}(\hat{\Z}_2^{o}\times \hat{\Z}_2^{v})$ SSB.} For this phase, we start from the GHZ Hamiltonian on the decoupled $o'$ qubits, and a diagonal symmetry breaking Hamiltonian on the chain formed by $o$ and $e$ qubits,
\beq
    \hat{H}_2=\sum_{j=1}^{L/2} -\hat{Z}'_{2j-1}\hat{Z}'_{2j+1}+\hat{Z}_{2j-1}\hat{Z}_{2j+1}-\hat{Y}_{2j}\hat{Y}_{2j+2}(1+\hat{Z}_{2j-1}\hat{Z}_{2j+3}).
\eeq
After gauging $\hat{\Z}_2^{o}\times\hat{\Z}_2^{o'}\times\hat{\Z}_2^{e}$, we obtain
\beq
    H_2=\sum_{j=1}^{L/2} -X'_{2j-1}+X_{2j-1}+Z_{2j-2}X_{2j}Z_{2j+2}(1+X_{2j-1}X_{2j+1}).
\eeq
The ground state is just given by
\beq
    \ket{\text{Rep}(G_{4,4}),2}=\ket{\text{Rep}(D_8),even}_{o,e}\otimes\ket{\text{product state}}_{o'}.
\eeq

\paragraph{3. Rep($G_{4,4}$) SPT from $\Z_2^4/\mathrm{diag}(\hat{\Z}_2^{o'}\times \hat{\Z}_2^{v})$ SSB.} This phase can be obtained from the last one by exchanging the $o$ and $o'$ qubits on the same sites. Therefore the SPT state is given by
\beq
    \ket{\text{Rep}(G_{4,4}),3}=\ket{\text{Rep}(D_8),even}_{o',e}\otimes\ket{\text{product state}}_{o}.
\eeq

\paragraph{4. Rep($G_{4,4}$) SPT from $\Z_2^4/\mathrm{diag}(\hat{\Z}_2^{e}\times \hat{\Z}_2^{v})$ SSB.} For this phase, we start from the following Hamiltonian,
\beq
    \hat{H}_4^{\pm}=\sum_{j=1}^{L/2} \pm\hat{Z}_{2j-1}\hat{Z}_{2j+1}\mp \hat{Z}'_{2j-1}\hat{Z}'_{2j+1}-\hat{Y}_{2j}\hat{Y}_{2j+2}(1+\hat{Z}_{2j-1}\hat{Z}'_{2j-1}\hat{Z}_{2j+3}\hat{Z}'_{2j+3}).
\eeq
It can be checked that the Hamiltonian commutes with the localized symmetry operators
\beq
    \hat{X}_{2j}C\hat{Z}_{2j,2j-1}C\hat{Z}'_{2j,2j-1}C\hat{Z}_{2j,2j+1}C\hat{Z}'_{2j,2j+1},
\eeq
of only $\mathrm{diag}(\hat{\Z}_2^{e}\times \hat{\Z}_2^{v})$ in the $\Z_2^4$ group. Furthermore, the two Hamiltonians $\hat{H}_4^+$ and $\hat{H}_4^-$ can be related by $\prod_j \hat{X}_{4j+1}\hat{X}'_{4j+1}$, which is a depth-1 symmetric circuit under $\Z_2^4$. Thus from above we conclude that both $\hat{H}_4^+$ and $\hat{H}_4^-$ describe the $\Z_2^4/\mathrm{diag}(\hat{\Z}_2^{e}\times \hat{\Z}_2^{v})$ SSB phase. After gauging $\hat{\Z}_2^{o}\times\hat{\Z}_2^{o'}\times\hat{\Z}_2^{e}$, we obtain
\beq
    H_4^{\pm}=\sum_{j=1}^{L/2} \pm X_{2j-1}\mp X'_{2j-1}+Z_{2j-2}X_{2j}Z_{2j+2}(1+X_{2j-1}X'_{2j-1}X_{2j+1}X'_{2j+1}).
\eeq

\paragraph{5. Rep($G_{4,4}$) SPT from $\Z_2^4/\mathrm{diag}(\hat{\Z}_2^{o}\times\hat{\Z}_2^{o'}\times \hat{\Z}_2^{v})$ SSB.} For this phase, we start from the following Hamiltonian
\beq
    \hat{H}_5=\sum_{j=1}^{L/2} \hat{Z}_{2j}\hat{Z}_{2j+2}-\hat{Y}_{2j-1}\hat{Y}_{2j+1}(1+\hat{Z}_{2j-2}\hat{Z}_{2j+2})-\hat{Y}'_{2j-1}\hat{Y}'_{2j+1}(1+\hat{Z}_{2j-2}\hat{Z}_{2j+2}),
\eeq
and gauge $\hat{\Z}_2^{o}\times\hat{\Z}_2^{o'}\times\hat{\Z}_2^{e}$. As s result we obtain
\beq
    H_5=\sum_{j=1}^{L/2}  X_{2j} + Z_{2j-3}X_{2j-1}Z_{2j+1}(1+X_{2j-2}X_{2j}) + Z'_{2j-3}X'_{2j-1}Z'_{2j+1}(1+X_{2j-2}X_{2j}).
\eeq


\paragraph{6. Rep($G_{4,4}$) SPT from $\Z_2^4/\mathrm{diag}(\hat{\Z}_2^4)$ SSB.} For this phase, we start from the following Hamiltonian
\beq
    \hat{H}_6^{\pm}=& \sum_{j=1}^{L/2} \pm \hat{Z}_{2j-1}\hat{Z}_{2j}\hat{Z}_{2j+1}\hat{Z}_{2j+2} \mp \hat{Z}'_{2j-1}\hat{Z}_{2j}\hat{Z}'_{2j+1}\hat{Z}_{2j+2} \\
    &- \hat{X}_{2j-1}\hat{X}'_{2j-1}\hat{Y}_{2j}\hat{X}_{2j+1}\hat{X}'_{2j+1}\hat{Y}_{2j+2}(1+\hat{Z}_{2j-1}\hat{Z}'_{2j-1}\hat{Z}_{2j+3}\hat{Z}'_{2j+3}).
\eeq
It is straightforward to check that $\hat{H}_2^{\pm}$ commute with commute with the localized symmetry operators
\beq
\hat{X}_{2j-1}\hat{X}'_{2j-1}\hat{X}_{2j}C\hat{Z}_{2j,2j-1}C\hat{Z}'_{2j,2j-1}C\hat{Z}_{2j,2j+1}C\hat{Z}'_{2j,2j+1}.
\eeq
of only the $\mathrm{diag}(\Z_2^4)$ subgroup of $\Z_2^4$. Again, the two Hamiltonians $\hat{H}_6^+$ and $\hat{H}_6^-$ are related by $\prod_j \hat{X}_{4j+1}\hat{X}'_{4j+1}$, which is a depth-1 symmetric circuit under $\Z_2^4$. Thus from above we conclude that both $\hat{H}_6^+$ and $\hat{H}_6^-$ describe the $\Z_2^4/\mathrm{diag}(\Z_2^4)$ SSB phase. Upon gauging $\hat{\Z}_2^{o}\times \hat{\Z}_2^{o'}\times \hat{\Z}_2^{e}$, $\hat{H}_6^+$ is mapped to
\beq
    H_6^{+}=&\sum_{j=1}^{L/2} X_{2j-1}X_{2j} - X'_{2j-1}X_{2j} \\
    &+ Z_{2j-3}Z'_{2j-3}Z_{2j-2}X_{2j}Z_{2j+1}Z'_{2j+1}Z_{2j+2}(1+X_{2j-1}X'_{2j-1}X_{2j+1}X'_{2j+1}).
\eeq

\subsection{General Rep($G$) symmetry}
We note that in the above examples, we consider the SPT states for Rep($D_8$), Rep($Q_8$), Rep($G_1$) and Rep($G_{4,4}$). The same procedure works for the representation category of any given class-2 nilpotent group $G$. Namely, for any such group $G$, we can always write down the symmetry operators for some anomalous abelian symmetry from the bulk SPT, such that upon gauging some subgroup, we can obtain the non-invertible Rep($G$) symmetry. The SPT phases one-to-one correspond to the possible symmetry breaking phases of abelian symmetry constrained by the anomaly. We further note that if the group $G$ is not class-2 nilpotent, but has a non-trivial center $Z$, there still exists such a duality between Rep($G$) non-invertible symmetry and $G/Z\times Z$ anomalous symmetry. This duality can be realized upon gauging $G/Z$ symmetry. Moreover, as in general the TY categories are always dual to some anomalous non-abelian symmetries~\cite{meir2012module,bhardwaj2018finite} (see also Appendix \ref{app:tachikawa}), it will be interesting to consider gauging some anomalous non-abelian symmetries on lattice to obtain the lattice models for SPT phases of general TY categories. We note that the anomalous symmetry operators can always be obtained from bulk SPT lattice models. However, the symmetry operators are not formed by local Clifford gates anymore for non-abelian groups. 

\section{Symmetry duality from bulk symmetry fractionalization}
\label{sec:SymFrac}
In this section, we first discuss the symmetry fractionalization in 2+1d SET phases. Then from the symmetry fractionalization in the bulk, we describe an understanding of the emergent non-abelian symmetries and non-invertible symmetries in 1+1d systems after gauging anomalous abelian symmetries.

The 2+1d SET (symmetry-enriched topological) phases, in general, are characterized by the anyon theory (UMTC), symmetry actions as automorphism on anyons, symmetry defectification, and symmetry fractionalization~\cite{barkeshli2019symmetry}.
All this structure combined forms a $G$-crossed braided tensor category. If we denote the symmetry group to be $G/N$, different symmetry defectification classes (SDC) can be obtained by stacking an extra $G/N$-SPTs on top of an SET. Thus the SDC form an $H^3(G/N,U(1))$ torsor. Different fractionalization classes form a $H^2(G/N,\mathcal{A})$ torsor, where $\mathcal{A}$ is the abelian group formed by fusions of abelian anyons.

\begin{figure}[h]
     \centering
     \begin{subfigure}[b]{0.3\textwidth}
         \centering
         \includegraphics[width=\textwidth]{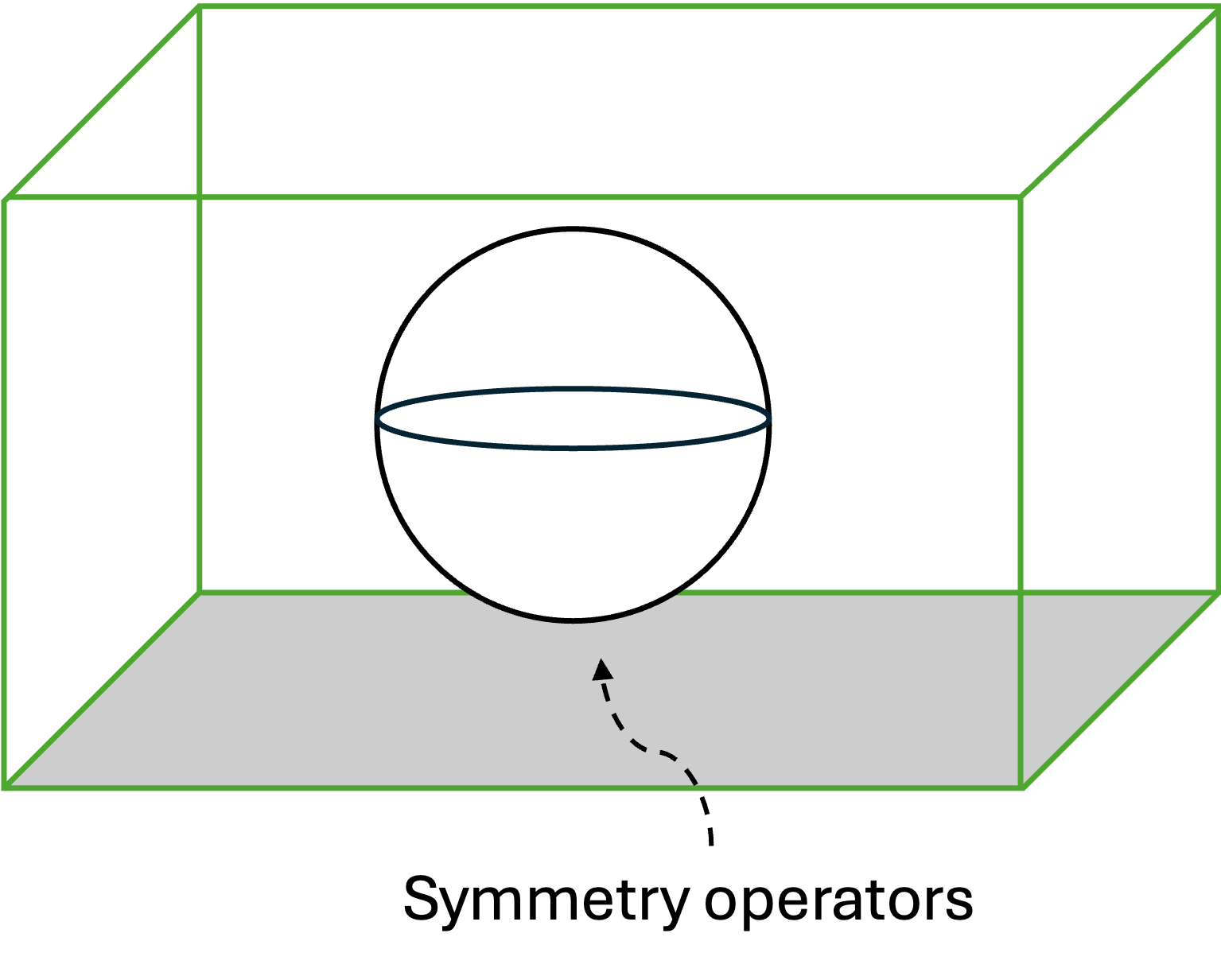}
         \caption{Bulk symmetry operators on codim-1 surfaces with group-like fusion rules.}
     \end{subfigure}
     \hfill
     \begin{subfigure}[b]{0.3\textwidth}
         \centering
         \includegraphics[width=\textwidth]{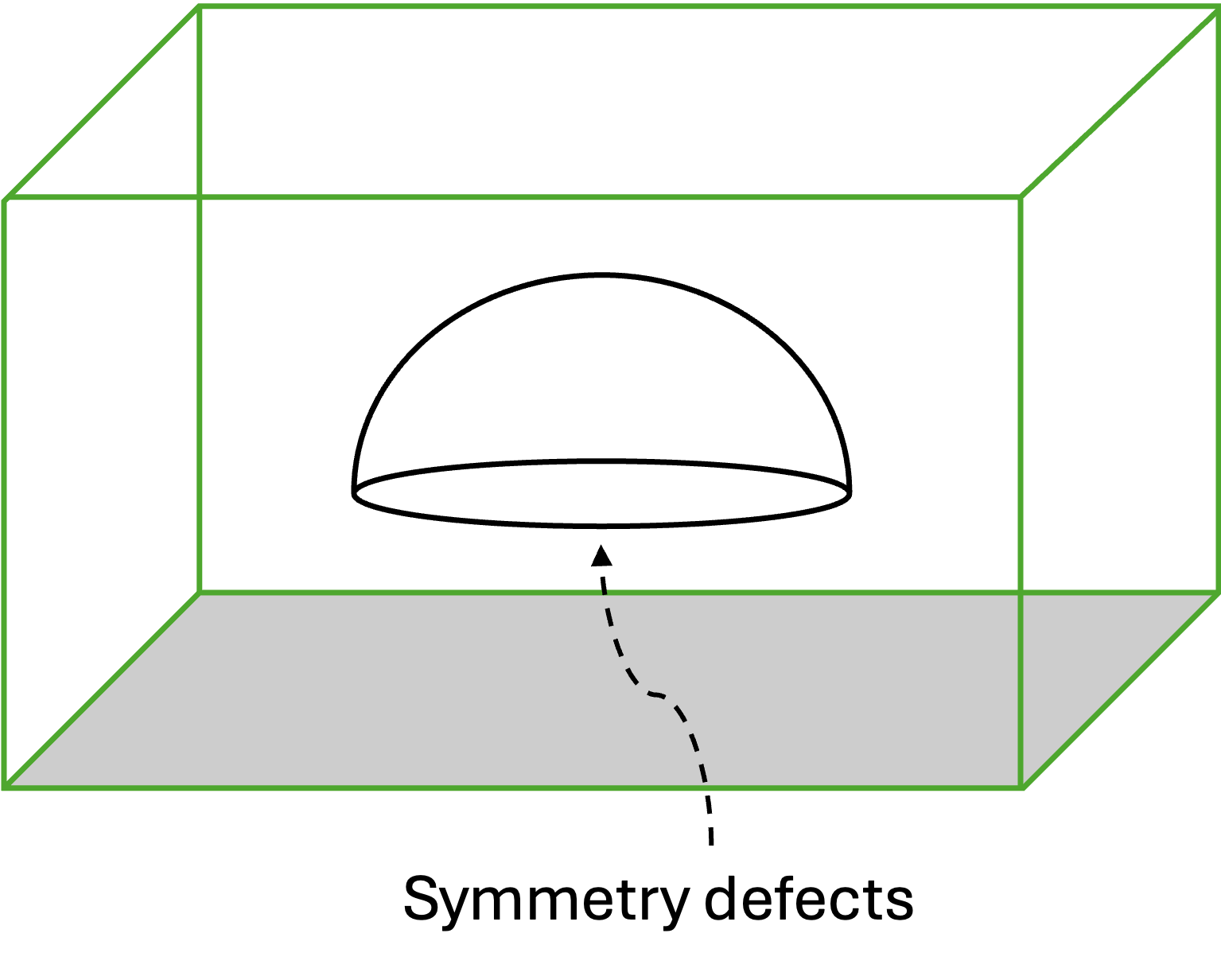}
         \caption{Bulk symmetry defects on the (codim-2) boundary of symmetry operators.}
     \end{subfigure}
     \hfill
     \begin{subfigure}[b]{0.3\textwidth}
         \centering
         \includegraphics[width=\textwidth]{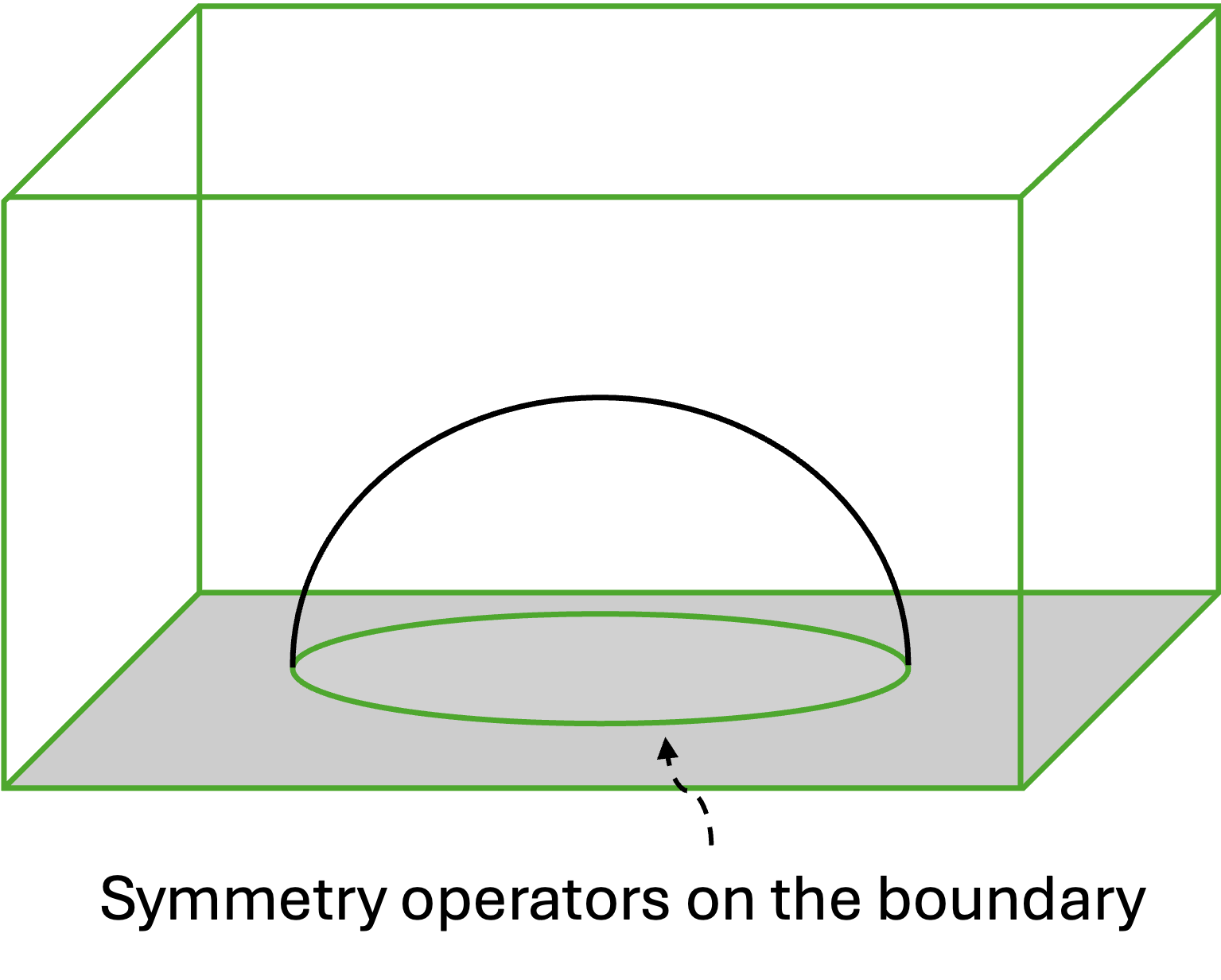}
         \caption{Boundary symmetry operators on codim-2 surfaces with richer fusion rules.}
     \end{subfigure}
        \caption{Bulk symmetry fractionalization describes the fusion of the bulk symmetry defects. It captures the fusion rules of symmetry operators on the boundary, which could become non-abelian or non-invertible.}
        \label{fig:sym_frac}
\end{figure}

From the last section, we saw from lattice models that, after gauging subgroups of certain anomalous symmetry, we might obtain non-abelian symmetry or non-invertible symmetry. Throughout this section, we will discuss the bulk counterpart, where after gauging a subgroup from an bulk SPT we obtain an SET phase with certain symmetry fractionalization. We will show that the fusion rules of symmetry defects (boundary of symmetry operators) can capture the boundary symmetry algebra as in Fig.~\ref{fig:sym_frac}.

We consider a 2+1d $G$-SPT, which is characterized by a 3-cocycle $[\nu]\in H^3(G,U(1))$. We assume the group $G$ is abelian, and we choose $N$ to be a subgroup of $G$. For our purpose, we only consider here the case when $\nu|_N\equiv 1$, such that the $N$ symmetry on the boundary is free of anomaly. Upon gauging a $N$ subgroup, we obtain an SET phase. The 0-form symmetry of this SET is the quotient group $G/N$. The anyon theory is just the quantum double of $N$ ($N$ toric code). 

The anyons in the $N$ toric code can be labeled by $(\rho,h)$, where $h\in N$ is the flux, and $\rho\in \hat{N}$ is the charge. For further calculation, we choose an embedding $s$ of quotient group $G/N$ into $G$. Now we consider a symmetry transformation labeled by $q\in G/N$. (we have a corresponding $G$ group element by the embedding map $s(q)$.)
The symmetry operator $U(q)$ will in general permute anyons as follows,
\beq
    (\rho,h)\xrightarrow[]{U(q)}(\rho',h'),
    \label{eq:anyon_permutation}
\eeq
where $\rho,\rho'\in\hat{N}$, and $h, h'\in N$.
Since we are considering abelian group $G$, the symmetry action on anyon theory is flux-preserving, thus $h'=h$. To fully describe the anyon permutation, we first define the slant product of the 3-cocycle $\nu$,
\beq
    c_h(g,k)\equiv \frac{\nu(h,g,k)\nu(g,k,h)}{\nu(g,h,k)}.
\eeq
One can check that $c_h$ is a 2-cocycle in $H^2(G,U(1))$ for any $h\in G$. Upon the symmetry action $U(q)$, anyon $(\rho, h)$ will be mapped to $(\rho',h)$, where the new charge $\rho'$ is
\beq
\rho'(g)=\frac{c_h(s(q),g)}{c_h(g,s(q))}\rho(g).
\label{eq:abelian_sym_automorphism}
\eeq

It is proved in Ref.~\cite{barkeshli2019symmetry} that, when a symmetry does not permute anyon types, the symmetry defects will always have fusion rules of the following form

\beq
    \eta_{q_1} \times \eta_{q_2} = \omega(q_1,q_2) \eta_{q_1 q_2},
\eeq
where $q_1,q_2$ are elements in the symmetry group $G/N$. And different cohomology classes of anyon $\omega$ in $H^2(G/N,\mathcal{A}=N\times \hat{N})$ characterize different symmetry fractionalization classes, thus different SET phases. Indeed, in our case the anyon $\omega(q_1,q_2)$ can be given from the 3-cocycle $\nu$ as
\beq
    \omega(q_1,q_2)=(\beta(q_1,q_2),s(q_1)s(q_2)s(q_1 q_2)^{-1}),
\eeq 
where the flux part of $\omega(q_1,q_2)$ depends on the embedding $s$, and the charge part $\beta$ is given by~\cite{li2023symmetry}
\beq
    \beta(q_1,q_2)(h)=\frac{\epsilon_{s(q_1)}(h)\epsilon_{s(q_2)}(h)}{\epsilon_{s(q_1q_1)}(h)}c_{h}(s(q_1),s(q_2)).
    \label{eq:defect_fusion}
\eeq
The $\epsilon_h(g)$ above is a 1-cochain such that its coboundary gives rise to the slant product $c_h(g,k)$,
\beq
    c_h(g,k)=\delta \epsilon_h(g,k)=\frac{\epsilon_h(k)\epsilon_h(g)}{\epsilon_h(gk)},
\eeq
for any $g,k\in N$.

\subsection{Non-abelian symmetries}
Now let us get back to the 1+1d problem of gauging a subgroup of an anomalous symmetry. 

We put the anomalous theory on the boundary of an SPT, and gauge the subgroup both in the bulk and on the boundary. On the boundary, we will have an emergent symmetry after gauging. In the bulk, there is an SET phase, with certain symmetry fractionalization as we just discussed. When the symmetry fractionalization of the bulk SET is non-trivial, i.e. $\omega(q_1,q_2)\neq 1$ for some $q_1,q_2\in G/N$, the symmetry could be extended after gauging. Further, when $\omega(q_1,q_2)\neq \omega(q_2,q_1)$ for some $q_1,q_2\in G/N$, the symmetry will be extended to some non-abelian group after gauging. As discussed in the last section, after gauging the $\hat{\Z}_2^v$ symmetry from $\hat{\Z}_2^e\times \hat{\Z}_2^o\times \hat{Z}_2^v$ type III anomaly, we obtain a $D_8$ symmetry. In the bulk, after gauging the $\hat{\Z}_2^v$ symmetry from type III SPT, we obtain a toric code with $\hat{\Z}_2^e\times \hat{\Z}_2^o$ symmetry~\cite{li2023symmetry}. The symmetry fractionalization of the phase is described by a cocycle $[w(g,h)]\in H^2(\hat{\Z}_2^e\times \hat{\Z}_2^o,\Z_2\times \hat{\Z}_2)$, where $\Z_2\times \hat{\Z}_2=<e,m>$ is the abelian group of anyon in toric code. We write an element $g\in\hat{\Z}_2^e\times \hat{\Z}_2^o$ as $g=(g_e,g_o)$, where $g_e, g_o=0,1$. The 2-cocycle is
\beq
    w(g,h)=\begin{cases}
    e, & g_e=h_o=1,\\
    1, & \text{others.}
    \end{cases}
\eeq
Therefore, the symmetry defects $\eta_e$, $\eta_o$ and $\eta_{eo}=\eta_{oe}$ have fusion rules
\beq
    \eta_o\times \eta_o=\eta_e\times \eta_e=1,\ \eta_o\times \eta_e=\eta_{eo},\ \eta_e\times \eta_o=e\cdot \eta_{eo},
\eeq
where $1$ and $e$ are the vacuum and $e$ particle. On the boundary of this SET, in addition to the $\hat{\Z}_2^e\times \hat{\Z}_2^o$ symmetry, there is also a reduced 0-form symmetry, the symmetry domain wall of which is the $e$ line on the boundary.
Therefore, the above fusion rules together with the anyon fusion rule $e\times e=1$ become the group multiplication rules of symmetry, which is given by
\beq
    \Tilde{\eta}_o^2=1,\ \Tilde{\eta}_e^2=1,\ \Tilde{V}^2=1,\ \Tilde{\eta}_o \Tilde{\eta}_e=\Tilde{V} \Tilde{\eta}_e \Tilde{\eta}_o.
    \label{eq:D_8_symmetry_fractionalization}
\eeq

It is straightforward to see that $\tilde{\eta}_e$, $\tilde{\eta}_o$ and $\tilde{V}$ in Eq.~\eqref{eq:D_8_symmetry_fractionalization} form a $D_8$ symmetry group with the following embedding:
\beq
    \tilde{\eta}_e = x, \quad \tilde{\eta}_o = xa, \quad \tilde{V} = a^2.
\eeq

Now let us consider the gauging of $\hat{\Z}_2^v$ symmetry from $\hat{\Z}_2^e\times \hat{\Z}_2^o\times \hat{\Z}_2^v$ type II+II+III anomaly given by,
\beq
    \frac{1}{2}A_e\cup A_e\cup A_v +\frac{1}{2}A_o\cup A_o\cup A_v+\frac{1}{2}A_e\cup A_o\cup A_v.
\eeq
After gauging $\hat{\Z}_2^v$ symmetry we obtain in the bulk a toric code with $\hat{\Z}_2^e\times \hat{\Z}_2^o$ symmetry and symmetry fractionalization~\cite{li2023symmetry},
\beq
    \eta_o\times \eta_o=e,\ \eta_e\times \eta_e=e,\ \eta_o\times \eta_e=\eta_{eo},\ \eta_e\times \eta_o=e\cdot \eta_{eo}.
\eeq
Therefore, on the boundary we obtain symmetry multiplication rules as
\beq
    \Tilde{\eta}_o^2=\Tilde{V},\ \Tilde{\eta}_e^2=\Tilde{V},\ \Tilde{V}^2=1,\ \Tilde{\eta}_o \Tilde{\eta}_e=\Tilde{V} \Tilde{\eta}_e \Tilde{\eta}_o,
\eeq
which is indeed a $Q_8$ group.

If we consider gauging $\hat{\Z}_2^o$ from the bulk type II+II+III SPT, the $\hat{\Z}_2^v\times \hat{\Z}_2^e$ symmetry fractionalization of the toric code is given by
\beq
    \check{\eta}_e\times \check{\eta}_e=1,\ \check{\eta}_v\times \check{\eta}_v=e,\ \check{\eta}_e\times \check{\eta}_v=\check{\eta}_{ev},\
    \check{\eta}_v\times\check{\eta}_e=e\cdot \check{\eta}_{ev},
\eeq
On the boundary, these fusion rules give rise to a $D_8$ symmetry as shown from the previous section in Eq.~\eqref{eq:D_8 from Q_8}. Besides the symmetry fractionalization, after gauging we still have a $\hat{\Z}_2^v\times \hat{\Z}_2^e$ SPT in the bulk inherited from the type II+II+III SPT, 
\beq
    \frac{1}{2}A_v\cup A_e\cup A_e.
    \label{eq:bulk SET defectification}
\eeq

It is also straightforward to show that from the two $\hat{\Z}_2^4$ anomalies we consider in the previous section, after gauging $\Z_2^v$ symmetry, the fusion rules of symmetry defects indeed form a Pauli group and a $G_{4,4}$ group.

\subsection{Non-invertible symmetries}
In this section, we discuss the cases when gauging a subgroup from anomalous symmetries gives rise to non-invertible symmetries.

In the above examples, the symmetries in the SET phases do not permute anyon types. However, it could happen that the charge $\rho'$ does not equal $\rho$ under some symmetry action $U(q)$ in Eq.~\eqref{eq:anyon_permutation}. In this case, we cannot find a 1d projective representation $\epsilon_h$ of $N$ such that the projective phase is $c_h$, therefore Eq.~\eqref{eq:defect_fusion} is not applicable. However, there does exist some higher dimensional projective representation $\epsilon_h$. As a result, the defect fusion becomes non-invertible in this case, and it is given by
\beq
    \eta_{q_1}\times \eta_{q_2}=\Big(\oplus_i N_i(q_1,q_2)(\rho_i,s(q_1)s(q_2)s(q_1 q_2)^{-1})\Big)\cdot \eta_{q_1q_2},
    \label{eq:non-invertible_eta}
\eeq
where $\rho_i$ represent all the chargeons, i.e. $\rho_i\in \hat{N}$, and the multiplicities $N_i$ are given by
\beq
    N_i(q_1,q_2)=\frac{1}{|N|}\sum_{h\in N}\mathrm{tr}(\epsilon_{s(q_1)}(h))\mathrm{tr}(\epsilon_{s(q_2)}(h))\mathrm{tr}(\epsilon_{s(q_1 q_2)}^*(h))\rho_i^*(h)c_h(s(q_1),s(q_2)).
    \label{eq:non-invertible_multiplicity}
\eeq

In the 1+1d theory accordingly, after gauging $N$ from an anomalous abelian $G$ symmetry, we will have an emergent non-invertible symmetry. We will show how the formula works through the next few examples.

We take $\hat{\Z}_2^e\times \hat{\Z}_2^o\times \hat{\Z}_2^v$ type III anomaly as our first example. After gauging $\hat{\Z}_2^e\times \hat{\Z}_2^o$, we have know that we would obtain a Rep($D_8$) symmetry. In the bulk, after gauging $\hat{\Z}_2^e\times \hat{\Z}_2^o$ from type III SPT, we obtain $\Z_2^e\times \Z_2^o$ toric code with a $\hat{\Z}_2^v$ symmetry. The slant product of type III cocycle will be given by
\beq
    c_h(g,k)=\begin{cases}
    -1, & h=g_e=k_o=1,\\
    1, & \text{others,}
    \end{cases}
\eeq
where we take $h\in \hat{Z}_2^v$ and $g,k\in \hat{\Z}_2^e\times \hat{\Z}_2^o\times \hat{\Z}_2^v$. From Eq.~\eqref{eq:abelian_sym_automorphism}, the $\hat{\Z}_2^v$ symmetry action on anyons is given by
\beq
    e_1, e_2\rightarrow e_1, e_2,\ m_1\rightarrow e_2 m_1,\ m_2\rightarrow e_1 m_2.
\eeq

Writing elements in $\hat{\Z}_2^e\times \hat{\Z}_2^o$ as $\{1,e,o,eo\}$, the irreducible projective representation of $\hat{\Z}_2^e\times \hat{\Z}_2^o$ for $c_1$ is given by
\beq
    \epsilon_1(1)=\begin{pmatrix}
        1 & 0\\
        0 & 1
    \end{pmatrix},\ \epsilon_1(e)=\begin{pmatrix}
        0 & 1\\
        1 & 0
    \end{pmatrix},\ \epsilon_1(o)=\begin{pmatrix}
        1 & 0\\
        0 & -1
    \end{pmatrix},\ \epsilon_1(eo)=\begin{pmatrix}
        0 & 1\\
        -1 & 0
    \end{pmatrix}.
    \label{eq:projective_rep}
\eeq
Using Eq.~\eqref{eq:non-invertible_multiplicity}, we obtain the defect fusion rule for the $\hat{\Z}_2^v$ symmetry,
\beq
    \eta_1 \times \eta_1= 1+e_1+e_2+e_1e_2.
    \label{eq:defect_fusion}
\eeq
The same result is also obtained in Sec.~XI(K) of Ref.~\cite{barkeshli2019symmetry}\footnote{We note that according to Ref.~\cite{barkeshli2019symmetry}, there is more than one defect in $\mathcal{C}_1$, the one we are showing here corresponds to the ``bare" defect.}. As a result of this fusion rule, we see that on the boundary after gauging, we have an emergent non-invertible symmetry, the fusion rule of which is given by
\beq
    U_e^2=U_o^2=1,\ \mathsf{D}^2=1+U_e+U_o+U_e U_o.
\eeq

We can also consider type II+II+III anomaly or any other type ...+III anomaly for $\hat{\Z}_2^e\times \hat{\Z}_2^o\times \hat{\Z}_2^v$ symmetry. After gauging $\hat{\Z}_2^e\times \hat{\Z}_2^o$ in the bulk, in general, we would obtain some abelian twisted quantum double model with $\hat{\Z}_2^v$ symmetry. The projective representation of $\hat{\Z}_2^e\times \hat{\Z}_2^o$ for the slant product is always the product of some 1d representation and the 2d representation in Eq.~\eqref{eq:projective_rep}. Therefore, the defect fusion rule is always given as in Eq.~\eqref{eq:defect_fusion}.

\section{Acknoledgments}
We thank Sahand Seifnashri for the comments on the draft.
We are also grateful to Brandon Rayhaun, Yichul Choi, and Meng Cheng for useful discussions. This work was supported by the National Science Foundation under Award No. PHY 2310614 and PHY 2210533.

\printbibliography

@article{li2023symmetry,
  title = {Symmetry-enriched topological order from partially gauging symmetry-protected topologically ordered states assisted by measurements},
  author = {Li, Yabo and Sukeno, Hiroki and Mana, Aswin Parayil and Nautrup, Hendrik Poulsen and Wei, Tzu-Chieh},
  journal = {Phys. Rev. B},
  volume = {108},
  issue = {11},
  pages = {115144},
  numpages = {37},
  year = {2023},
  month = {Sep},
  publisher = {American Physical Society},
  doi = {10.1103/PhysRevB.108.115144},
  url = {https://link.aps.org/doi/10.1103/PhysRevB.108.115144}
}

@article{seifnashri2024cluster,
  title={Cluster state as a non-invertible symmetry protected topological phase},
  author={Seifnashri, Sahand and Shao, Shu-Heng},
  journal={arXiv preprint arXiv:2404.01369},
  year={2024},
  url = {https://arxiv.org/abs/2404.01369}
}

@article{hu2020electric,
  title={Electric-magnetic duality in twisted quantum double model of topological orders},
  author={Hu, Yuting and Wan, Yidun},
  journal={Journal of High Energy Physics},
  volume={2020},
  number={11},
  pages={1--39},
  year={2020},
  publisher={Springer},
  url={https://link.springer.com/article/10.1007/JHEP11(2020)170}
}

@article{propitius1995topological,
  title={Topological interactions in broken gauge theories},
  author={Propitius, Mark de Wild},
  journal={arXiv preprint hep-th/9511195},
  year={1995},
  url={https://arxiv.org/abs/hep-th/9511195}
}

@article{yoshida2017gapped,
title = {Gapped boundaries, group cohomology and fault-tolerant logical gates},
journal = {Annals of Physics},
volume = {377},
pages = {387-413},
year = {2017},
issn = {0003-4916},
doi = {https://doi.org/10.1016/j.aop.2016.12.014},
url = {https://www.sciencedirect.com/science/article/pii/S0003491616302858},
author = {Beni Yoshida}
}

@article{thorngren2019fusion,
  title={Fusion category symmetry I: anomaly in-flow and gapped phases},
  author={Thorngren, Ryan and Wang, Yifan},
  journal={arXiv preprint arXiv:1912.02817},
  year={2019},
  url={https://arxiv.org/abs/1912.02817}
}

@article{uribe2017classification,
  title={On the classification of pointed fusion categories up to weak Morita equivalence},
  author={Uribe Jongbloed, Bernardo},
  journal={Pacific Journal of Mathematics},
  volume={290},
  number={2},
  pages={437--466},
  year={2017},
  publisher={Mathematical Sciences Publishers},
  url={https://msp.org/pjm/2017/290-2/p08.xhtml}
}

@article{seifnashri2023lieb,
  title={Lieb-Schultz-Mattis anomalies as obstructions to gauging (non-on-site) symmetries},
  author={Seifnashri, Sahand},
  journal={arXiv preprint arXiv:2308.05151},
  year={2023},
  url={https://arxiv.org/abs/2308.05151}
}

@article{barkeshli2019symmetry,
  title={Symmetry fractionalization, defects, and gauging of topological phases},
  author={Barkeshli, Maissam and Bonderson, Parsa and Cheng, Meng and Wang, Zhenghan},
  journal={Physical Review B},
  volume={100},
  number={11},
  pages={115147},
  year={2019},
  publisher={APS},
  url={https://arxiv.org/abs/1410.4540}
}

@article{bhardwaj2018finite,
  title={On finite symmetries and their gauging in two dimensions},
  author={Bhardwaj, Lakshya and Tachikawa, Yuji},
  journal={Journal of High Energy Physics},
  volume={2018},
  number={3},
  pages={1--68},
  year={2018},
  publisher={Springer},
  url={https://link.springer.com/article/10.1007/JHEP03(2018)189}
}

@article{meir2012module,
  title={Module categories over graded fusion categories},
  author={Meir, Ehud and Musicantov, Evgeny},
  journal={Journal of Pure and Applied Algebra},
  volume={216},
  number={11},
  pages={2449--2466},
  year={2012},
  publisher={Elsevier},
  url={https://arxiv.org/abs/1010.4333}
}

@article{wang2015non,
  title={Non-Abelian string and particle braiding in topological order: Modular SL (3, Z) representation and (3+1)-dimensional twisted gauge theory},
  author={Wang, Juven C and Wen, Xiao-Gang},
  journal={Physical Review B},
  volume={91},
  number={3},
  pages={035134},
  year={2015},
  publisher={APS},
  url={https://arxiv.org/abs/1404.7854}
}

@article{levin2012braiding,
  title = {Braiding statistics approach to symmetry-protected topological phases},
  author = {Levin, Michael and Gu, Zheng-Cheng},
  journal = {Phys. Rev. B},
  volume = {86},
  issue = {11},
  pages = {115109},
  numpages = {15},
  year = {2012},
  month = {Sep},
  publisher = {American Physical Society},
  doi = {10.1103/PhysRevB.86.115109},
  url ={https://link.aps.org/doi/10.1103/PhysRevB.86.115109}
}

@article{kawagoe2021anomalies,
  title = {Anomalies in bosonic symmetry-protected topological edge theories: Connection to $F$ symbols and a method of calculation},
  author = {Kawagoe, Kyle and Levin, Michael},
  journal = {Phys. Rev. B},
  volume = {104},
  issue = {11},
  pages = {115156},
  numpages = {20},
  year = {2021},
  month = {Sep},
  publisher = {American Physical Society},
  doi = {10.1103/PhysRevB.104.115156},
  url = {https://link.aps.org/doi/10.1103/PhysRevB.104.115156}
}

@article{diatlyk2024gauging,
  title={Gauging non-invertible symmetries: topological interfaces and generalized orbifold groupoid in 2d QFT},
  author={Diatlyk, Oleksandr and Luo, Conghuan and Wang, Yifan and Weller, Quinten},
  journal={Journal of High Energy Physics},
  volume={2024},
  number={3},
  pages={1--52},
  year={2024},
  publisher={Springer},
  url={https://arxiv.org/abs/2311.17044}
}

@article{seiberg2024majorana,
  title={Majorana chain and Ising model-(non-invertible) translations, anomalies, and emanant symmetries},
  author={Seiberg, Nathan and Shao, Shu-Heng},
  journal={SciPost Physics},
  volume={16},
  number={3},
  pages={064},
  year={2024},
  url={https://arxiv.org/abs/2307.02534}
}

@article{tambara2000representations,
  title={Representations of tensor categories with fusion rules of self-duality for abelian groups},
  author={Tambara, Daisuke},
  journal={Israel Journal of Mathematics},
  volume={118},
  pages={29--60},
  year={2000},
  publisher={Springer},
  url={https://link.springer.com/article/10.1007/BF02803515#citeas}
}

@article{chen2013symmetry,
   title = {Symmetry protected topological orders and the group cohomology of their symmetry group},
  author = {Chen, Xie and Gu, Zheng-Cheng and Liu, Zheng-Xin and Wen, Xiao-Gang},
  journal = {Phys. Rev. B},
  volume = {87},
  issue = {15},
  pages = {155114},
  numpages = {48},
  year = {2013},
  month = {Apr},
  publisher = {American Physical Society},
  doi = {10.1103/PhysRevB.87.155114},
  url = {https://link.aps.org/doi/10.1103/PhysRevB.87.155114}
}

@article{pollman2012aymmetry,
  title = {Symmetry protection of topological phases in one-dimensional quantum spin systems},
  author = {Pollmann, Frank and Berg, Erez and Turner, Ari M. and Oshikawa, Masaki},
  journal = {Phys. Rev. B},
  volume = {85},
  issue = {7},
  pages = {075125},
  numpages = {9},
  year = {2012},
  month = {Feb},
  publisher = {American Physical Society},
  doi = {10.1103/PhysRevB.85.075125},
  url = {https://link.aps.org/doi/10.1103/PhysRevB.85.075125}
}

@article{schuch2011classifying,
  title = {Classifying quantum phases using matrix product states and projected entangled pair states},
  author = {Schuch, Norbert and P\'erez-Garc\'{\i}a, David and Cirac, Ignacio},
  journal = {Phys. Rev. B},
  volume = {84},
  issue = {16},
  pages = {165139},
  numpages = {21},
  year = {2011},
  month = {Oct},
  publisher = {American Physical Society},
  doi = {10.1103/PhysRevB.84.165139},
  url = {https://link.aps.org/doi/10.1103/PhysRevB.84.165139}
}

@article{bhardwaj2024illustrating,
  title={Illustrating the Categorical Landau Paradigm in Lattice Models},
  author={Bhardwaj, Lakshya and Bottini, Lea E and Schafer-Nameki, Sakura and Tiwari, Apoorv},
  journal={arXiv preprint arXiv:2405.05302},
  year={2024},
  url={https://arxiv.org/abs/2405.05302}
}

@article{chen2011classification,
  title = {Classification of gapped symmetric phases in one-dimensional spin systems},
  author = {Chen, Xie and Gu, Zheng-Cheng and Wen, Xiao-Gang},
  journal = {Phys. Rev. B},
  volume = {83},
  issue = {3},
  pages = {035107},
  numpages = {19},
  year = {2011},
  month = {Jan},
  publisher = {American Physical Society},
  doi = {10.1103/PhysRevB.83.035107},
  url = {https://link.aps.org/doi/10.1103/PhysRevB.83.035107}
}

@article{turner2011topological,
  title = {Topological phases of one-dimensional fermions: An entanglement point of view},
  author = {Turner, Ari M. and Pollmann, Frank and Berg, Erez},
  journal = {Phys. Rev. B},
  volume = {83},
  issue = {7},
  pages = {075102},
  numpages = {11},
  year = {2011},
  month = {Feb},
  publisher = {American Physical Society},
  doi = {10.1103/PhysRevB.83.075102},
  url = {https://link.aps.org/doi/10.1103/PhysRevB.83.075102}
}

@article{lu2012theory,
  title = {Theory and classification of interacting integer topological phases in two dimensions: A Chern-Simons approach},
  author = {Lu, Yuan-Ming and Vishwanath, Ashvin},
  journal = {Phys. Rev. B},
  volume = {86},
  issue = {12},
  pages = {125119},
  numpages = {28},
  year = {2012},
  month = {Sep},
  publisher = {American Physical Society},
  doi = {10.1103/PhysRevB.86.125119},
  url = {https://link.aps.org/doi/10.1103/PhysRevB.86.125119}
}

@article{chen2014symmetry,
  title={Symmetry-protected topological phases from decorated domain walls},
  author={Chen, Xie and Lu, Yuan-Ming and Vishwanath, Ashvin},
  journal={Nature communications},
  volume={5},
  number={1},
  pages={3507},
  year={2014},
  publisher={Nature Publishing Group UK London},
  url={https://www.nature.com/articles/ncomms4507}
}

@article{li2023measuring,
  title={Measuring Topological Field Theories: Lattice Models and Field-Theoretic Description},
  author={Li, Yabo and Litvinov, Mikhail and Wei, Tzu-Chieh},
  journal={arXiv preprint arXiv:2310.17740},
  year={2023},
  url={https://arxiv.org/abs/2310.17740}
}

@article{senthil2015symmetry,
  title={Symmetry-protected topological phases of quantum matter},
  author={Senthil, Todadri},
  journal={Annu. Rev. Condens. Matter Phys.},
  volume={6},
  number={1},
  pages={299--324},
  year={2015},
  publisher={Annual Reviews},
  url={https://www.annualreviews.org/content/journals/10.1146/annurev-conmatphys-031214-014740}
}

@article{briegel2009measurement,
  title={Measurement-based quantum computation},
  author={Briegel, Hans J and Browne, David E and D{\"u}r, Wolfgang and Raussendorf, Robert and Van den Nest, Maarten},
  journal={Nature Physics},
  volume={5},
  number={1},
  pages={19--26},
  year={2009},
  publisher={Nature Publishing Group UK London},
  url={https://www.nature.com/articles/nphys1157}
}

@article{raussendorf2002one,
  title={The one-way quantum computer--a non-network model of quantum computation},
  author={Raussendorf, Robert and Browne, Daniel and Briegel, Hans},
  journal={journal of modern optics},
  volume={49},
  number={8},
  pages={1299--1306},
  year={2002},
  publisher={Taylor \& Francis},
  url={https://www.tandfonline.com/doi/abs/10.1080/09500340110107487}
}

@article{miyake2010quantum,
  title = {Quantum Computation on the Edge of a Symmetry-Protected Topological Order},
  author = {Miyake, Akimasa},
  journal = {Phys. Rev. Lett.},
  volume = {105},
  issue = {4},
  pages = {040501},
  numpages = {4},
  year = {2010},
  month = {Jul},
  publisher = {American Physical Society},
  doi = {10.1103/PhysRevLett.105.040501},
  url = {https://link.aps.org/doi/10.1103/PhysRevLett.105.040501}
}

@article{else2012symmetry,
  title = {Symmetry-Protected Phases for Measurement-Based Quantum Computation},
  author = {Else, Dominic V. and Schwarz, Ilai and Bartlett, Stephen D. and Doherty, Andrew C.},
  journal = {Phys. Rev. Lett.},
  volume = {108},
  issue = {24},
  pages = {240505},
  numpages = {5},
  year = {2012},
  month = {Jun},
  publisher = {American Physical Society},
  doi = {10.1103/PhysRevLett.108.240505},
  url = {https://link.aps.org/doi/10.1103/PhysRevLett.108.240505}
}

@article{stephen2017computational,
  title = {Computational Power of Symmetry-Protected Topological Phases},
  author = {Stephen, David T. and Wang, Dong-Sheng and Prakash, Abhishodh and Wei, Tzu-Chieh and Raussendorf, Robert},
  journal = {Phys. Rev. Lett.},
  volume = {119},
  issue = {1},
  pages = {010504},
  numpages = {5},
  year = {2017},
  month = {Jul},
  publisher = {American Physical Society},
  doi = {10.1103/PhysRevLett.119.010504},
  url = {https://link.aps.org/doi/10.1103/PhysRevLett.119.010504}
}

@article{raussendorf2017symmetry,
  title = {Symmetry-protected topological phases with uniform computational power in one dimension},
  author = {Raussendorf, Robert and Wang, Dong-Sheng and Prakash, Abhishodh and Wei, Tzu-Chieh and Stephen, David T.},
  journal = {Phys. Rev. A},
  volume = {96},
  issue = {1},
  pages = {012302},
  numpages = {14},
  year = {2017},
  month = {Jul},
  publisher = {American Physical Society},
  doi = {10.1103/PhysRevA.96.012302},
  url = {https://link.aps.org/doi/10.1103/PhysRevA.96.012302}
}

@article{wei2017universal,
  title = {Universal measurement-based quantum computation in two-dimensional symmetry-protected topological phases},
  author = {Wei, Tzu-Chieh and Huang, Ching-Yu},
  journal = {Phys. Rev. A},
  volume = {96},
  issue = {3},
  pages = {032317},
  numpages = {12},
  year = {2017},
  month = {Sep},
  publisher = {American Physical Society},
  doi = {10.1103/PhysRevA.96.032317},
  url = {https://link.aps.org/doi/10.1103/PhysRevA.96.032317}
}

@article{wei2018quantum,
  title={Quantum spin models for measurement-based quantum computation},
  author={Wei, Tzu-Chieh},
  journal={Advances in Physics: X},
  volume={3},
  number={1},
  pages={1461026},
  year={2018},
  publisher={Taylor \& Francis},
  url={https://www.tandfonline.com/doi/full/10.1080/23746149.2018.1461026}
}

@article{raussendorf2019computationally,
  title = {Computationally Universal Phase of Quantum Matter},
  author = {Raussendorf, Robert and Okay, Cihan and Wang, Dong-Sheng and Stephen, David T. and Nautrup, Hendrik Poulsen},
  journal = {Phys. Rev. Lett.},
  volume = {122},
  issue = {9},
  pages = {090501},
  numpages = {5},
  year = {2019},
  month = {Mar},
  publisher = {American Physical Society},
  doi = {10.1103/PhysRevLett.122.090501},
  url = {https://link.aps.org/doi/10.1103/PhysRevLett.122.090501}
}

@article{fechisin2023non,
  title={Non-invertible symmetry-protected topological order in a group-based cluster state},
  author={Fechisin, Christopher and Tantivasadakarn, Nathanan and Albert, Victor V},
  journal={arXiv preprint arXiv:2312.09272},
  year={2023},
  url={https://arxiv.org/abs/2312.09272}
}

@article{bhardwaj2024lattice,
  title={Lattice Models for Phases and Transitions with Non-Invertible Symmetries},
  author={Bhardwaj, Lakshya and Bottini, Lea E and Schafer-Nameki, Sakura and Tiwari, Apoorv},
  journal={arXiv preprint arXiv:2405.05964},
  year={2024},
  url={https://arxiv.org/abs/2405.05964}
}

@article{inamura2021topological,
  title={Topological field theories and symmetry protected topological phases with fusion category symmetries},
  author={Inamura, Kansei},
  journal={Journal of High Energy Physics},
  volume={2021},
  number={5},
  pages={1--35},
  year={2021},
  publisher={Springer},
  url={https://arxiv.org/abs/2103.15588}
}

@article{mcgreevy2023generalized,
  title={Generalized symmetries in condensed matter},
  author={McGreevy, John},
  journal={Annual Review of Condensed Matter Physics},
  volume={14},
  pages={57--82},
  year={2023},
  publisher={Annual Reviews},
  url={https://www.annualreviews.org/content/journals/10.1146/annurev-conmatphys-040721-021029}
}

@article{cordova2022snowmass,
  title={Snowmass white paper: Generalized symmetries in quantum field theory and beyond},
  author={Cordova, Clay and Dumitrescu, Thomas T and Intriligator, Kenneth and Shao, Shu-Heng},
  journal={arXiv preprint arXiv:2205.09545},
  year={2022},
  url={https://arxiv.org/abs/2205.09545}
}

@article{brennan2023introduction,
  title={Introduction to generalized global symmetries in QFT and particle physics},
  author={Brennan, T Daniel and Hong, Sungwoo},
  journal={arXiv preprint arXiv:2306.00912},
  year={2023},
  url={https://arxiv.org/abs/2306.00912}
}

@article{schafer2024ictp,
  title={ICTP lectures on (non-) invertible generalized symmetries},
  author={Sch{\"a}fer-Nameki, Sakura},
  journal={Physics Reports},
  volume={1063},
  pages={1--55},
  year={2024},
  publisher={Elsevier},
  url={https://arxiv.org/abs/2305.18296}
}

@article{shao2023s,
  title={What's Done Cannot Be Undone: TASI Lectures on Non-Invertible Symmetry},
  author={Shao, Shu-Heng},
  journal={arXiv preprint arXiv:2308.00747},
  year={2023},
  url={https://arxiv.org/abs/2308.00747}
}

@article{kitaev2012models,
  title={Models for gapped boundaries and domain walls},
  author={Kitaev, Alexei and Kong, Liang},
  journal={Communications in Mathematical Physics},
  volume={313},
  number={2},
  pages={351--373},
  year={2012},
  publisher={Springer},
  url={https://link.springer.com/article/10.1007/s00220-012-1500-5}
}

\numberwithin{equation}{section}
\appendix
\section{Partial electric-magnetic (PEM) duality}
\label{app:PEM}
In this section, we review the conditions for the PEM duality in 2+1d and show some examples when abelian twisted quantum double models are dual to certain non-abelian quantum double models. 

Assume that symmetry groups are $G=N\ltimes_F K$, and $G=\hat{N}\ltimes_{\hat{F}} K$, where $N$ is a normal abelian subgroup, and $\hat{N}$ is the Pontryagin dual, and $F\in H^2(K,N)$.
There exists an identification between two twisted quantum double $D^{\alpha}(G)$ and $D^{\alpha'}(G')$ under certain conditions, by exchanging the $N$-charges and $N$-fluxes. This is called a PEM (partial electric-magnetic) duality~\cite{uribe2017classification,hu2020electric}.
The condition on those two topological orders to be equivalent: there exists a 3-cochain $\epsilon \in C^3(K, U(1))$, such that $\delta_K \epsilon=\hat{F} \wedge F$, i.e.,
\begin{equation}
\delta_K \epsilon\left(k_1, k_2, k_3, k_4\right)=\frac{\epsilon\left(k_2, k_3, k_4\right) \epsilon\left(k_1, k_2 k_3, k_4\right) \epsilon\left(k_1, k_2, k_3\right)}{\epsilon\left(k_1 k_2, k_3, k_4\right) \epsilon\left(k_1, k_2, k_3 k_4\right)}=\hat{F}\left(k_1, k_2\right)\left(F\left(k_3, k_4\right)\right)
\end{equation}

Assuming that the above conditions is satisfied there is a weakly Morita equivalence $\mathrm{Vec}_G^\alpha \cong \mathrm{Vec}_{G^{\prime}}^{\alpha^{\prime}}$, with $G=N \ltimes_F K, G^{\prime}= \hat{N}\ltimes_{\hat{F}} K$, and the 3 -cocycles are 
\begin{equation}
\begin{aligned}
\alpha\left(\left(a_1, k_1\right),\left(a_2, k_2\right),\left(a_3, k_3\right)\right) & =\hat{F}\left(k_1, k_2\right)\left(a_3\right) \epsilon\left(k_1, k_2, k_3\right) \\
\alpha^{\prime}\left(\left(x_1, \rho_1\right),\left(x_2, \rho_2\right),\left(x_3, \rho_3\right)\right) & =\rho_1\left(F\left(x_2, x_3\right)\right) \epsilon\left(x_1, x_2, x_3\right)
\end{aligned}
\label{eq:PEM equation}
\end{equation}

Taking $N=\Z_2$ and $K=\Z_2\times \Z_2$, the results of group extension can be $D_8$, $Q_8$, $\Z_4\times \Z_2$ and $\Z_2^3$. In fact from PEM duality, the quantum double of all these groups are identical to some twisted quantum double of $\Z_2^3$. Here we will show the PEM for $D_8$ and $Q_8$ group.
The group extensions give rise to two non-abelian groups, $D_8$ and $Q_8$. 

Throughout the paper we will use topological action to represent cocycles of abelian groups. For example, a type III topological action\footnote{More details about different types of cocycles are in Appendix \ref{app:Z_2^3}.}
\beq
    \alpha=\frac{1}{2}A_1\cup A_2\cup A_3,
\eeq
corresponds to a 3-cocycle as
\beq
    \alpha(g,h,k)=e^{\frac{2\pi i }{2}\delta_{g_1,1}\delta_{h_2,1}\delta_{k_3,1}}.
\eeq
We denote an element $g\in \Z_2^3$ as $g=(g_1,g_2,g_3)$, where $g_1,g_2,g_3=0,1$. If we take $N$ to be the third $\Z_2$ group, since $F\equiv0$, $\hat{F}\wedge F\equiv 0$. From Eq.~\eqref{eq:PEM equation}, 
\beq
    \hat{F}=\frac{1}{2}A_1\cup A_2.
\eeq
From $[\hat{F}]\in H^2(\Z_2^2,\hat{\Z}_2)$, we can define a group extension. It is straightforward to check that this extended group is $D_8$.

If we start from a type II+II+III $\Z_2^3$ twisted quantum double
\beq
    \alpha=\frac{1}{2}A_1\cup A_1\cup A_3+\frac{1}{2}A_2\cup A_2\cup A_3+\frac{1}{2}A_1\cup A_2\cup A_3,
\eeq
and again take $N$ to be the third $\Z_2$ group, according to Eq.~\eqref{eq:PEM equation}, the 2-cocycle for the group extension on the dual side is given by
\beq
    \hat{F}=\frac{1}{2}A_1\cup A_1+\frac{1}{2}A_2\cup A_2+\frac{1}{2}A_1\cup A_2.
\eeq
This extension gives rise to a $Q_8$ group.

If we start from $N=\Z_2$, $K=\Z_2^3$ and $G=K\times N$, the derivation is very similar. For the $\Z_2^4$ 3-cocycle
\beq
    \alpha=\frac{1}{2}A_1\cup A_2\cup A_4 + \frac{1}{2}A_3\cup A_3\cup A_4,
\eeq    
the dual side should have an extension of $\Z_2^3$ by $\Z_2$ given by a 2-cocycle
\beq
    \hat{F}=\frac{1}{2}A_1\cup A_2 + \frac{1}{2}A_3\cup A_3,
\eeq
which is a Pauli group.

Alternatively, for the $\Z_2^4$ 3-cocycle
\beq
    \alpha=\frac{1}{2}A_1\cup A_3\cup A_4 + \frac{1}{2}A_2\cup A_3\cup A_4,
\eeq    
the dual gauge group given by 2-cocycle
\beq
    \hat{F}=\frac{1}{2}A_1\cup A_3 + \frac{1}{2}A_2\cup A_3,
\eeq
is called group $G_{4,4}$.

We have two comments here. First, for the quantum double of any class-2 nilpotent group $G$, it is always equivalent to the twisted quantum double of an abelian group. Second, the result for $Q_8$ quantum double above is falsifying one of the claims in Ref.~\cite{propitius1995topological}, where the $Q_8$ quantum double is claimed to be dual to a $\Z_2^3$ twisted quantum double with a type I+III cocycle.

\section{2+1d $\Z_2^3$ SPT lattice models}
\label{app:Z_2^3}
In this section, we construct the lattice models for all types of 2+1d $Z_2^3$ SPT by performing symmetry breaking from the lattice model of type-III SPT and stacking.

The 2+1d SPT phases under $\Z_2^3$ symmetry is classified by $H^3(\Z_2^3,U(1))=\Z_2^7$, in which the root states can be characterized as three types~\cite{propitius1995topological,wang2015non,li2023measuring}.
Type I SPT is a non-trivial SPT only under one of the $\Z_2$ groups, the topological action of which is given by $\frac{1}{2}A\cup A\cup A$. Type II SPT is a non-trivial SPT under a $\Z_2^2$ subgroup. Their topological action is given by $\frac{1}{2}A_1\cup A_2\cup A_2$. And type III SPT has topological action $\frac{1}{2}A_1\cup A_2\cup A_3$. Therefore, suppose we have a type III SPT state under symmetry $\Z_2^e\times \Z_2^o\times\Z_2^v$, by breaking the symmetry into $\Z_2^v\times \mathrm{diag}(\Z_2^e\times \Z_2^o)$, we would obtain a type II SPT state. Further breaking the symmetry into $\mathrm{diag}(\Z_2^3)$, we would obtain a type I SPT.

\begin{figure}[h]
    \centering
    \includegraphics[width=0.8\linewidth]{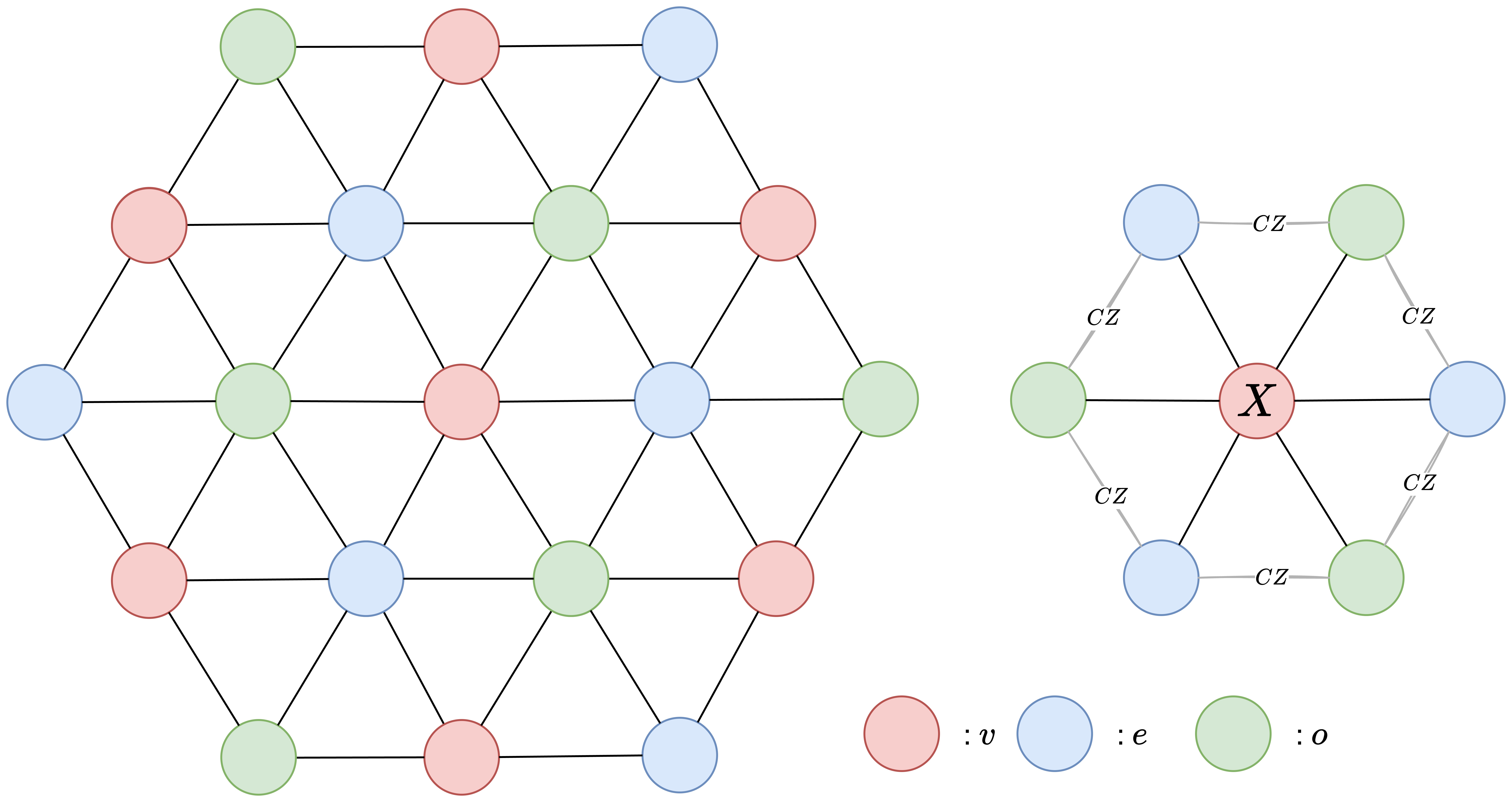}
    \caption{The 3-colorable lattice and the type-III SPT stabilizers, }
    \label{fig:type-III}
\end{figure}

The type-III SPT state can be given on 3-colorable triangular lattice by applying $CCZ$ (controlled-controlled-$Z$) gate between qubits on $v,e,o$ sites, when they are the vertices of a same triangle~\cite{yoshida2017gapped}. For the convenience of later discussion, we use $v,e,o$ to label three colors. The symmetry operators are product of Pauli $X$ operators on sites of each color,
\beq
    U_v=\prod_v X_v,\ U_e=\prod_e X_e,\ U_o=\prod_o X_o.
\eeq
And the stabilizers of the SPT states are given in Fig.~\ref{fig:type-III}.

\begin{figure}[h]
    \centering
    \includegraphics[width=0.9\linewidth]{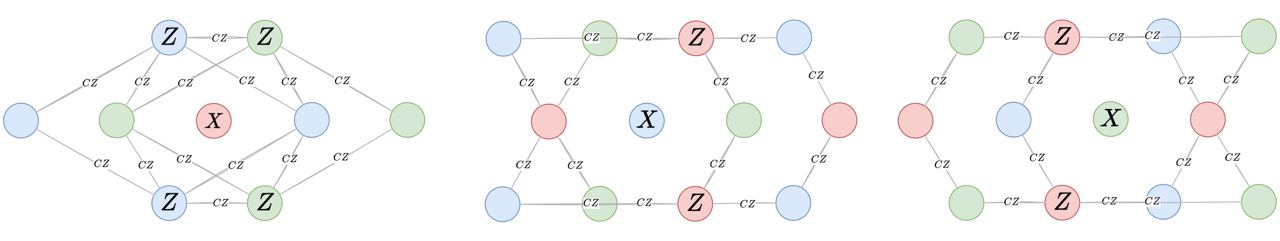}
    \caption{Stabilizers for type II+II+III SPT given by topological action $\frac{1}{2}A_e\cup A_e\cup A_v+\frac{1}{2}A_o\cup A_o\cup A_v+\frac{1}{2}A_e\cup A_o\cup A_v$.}
    \label{fig:type-II+II+III}
\end{figure}

Now to break the $Z_2^e\times Z_2^o$ symmetry into a diagonal subgroup, we add a hopping term in the Hamiltonian,
\beq
    H_{coupling}=\vcenter{\hbox{\includegraphics[scale=.03,trim={0cm 0cm 0cm 0cm},clip]{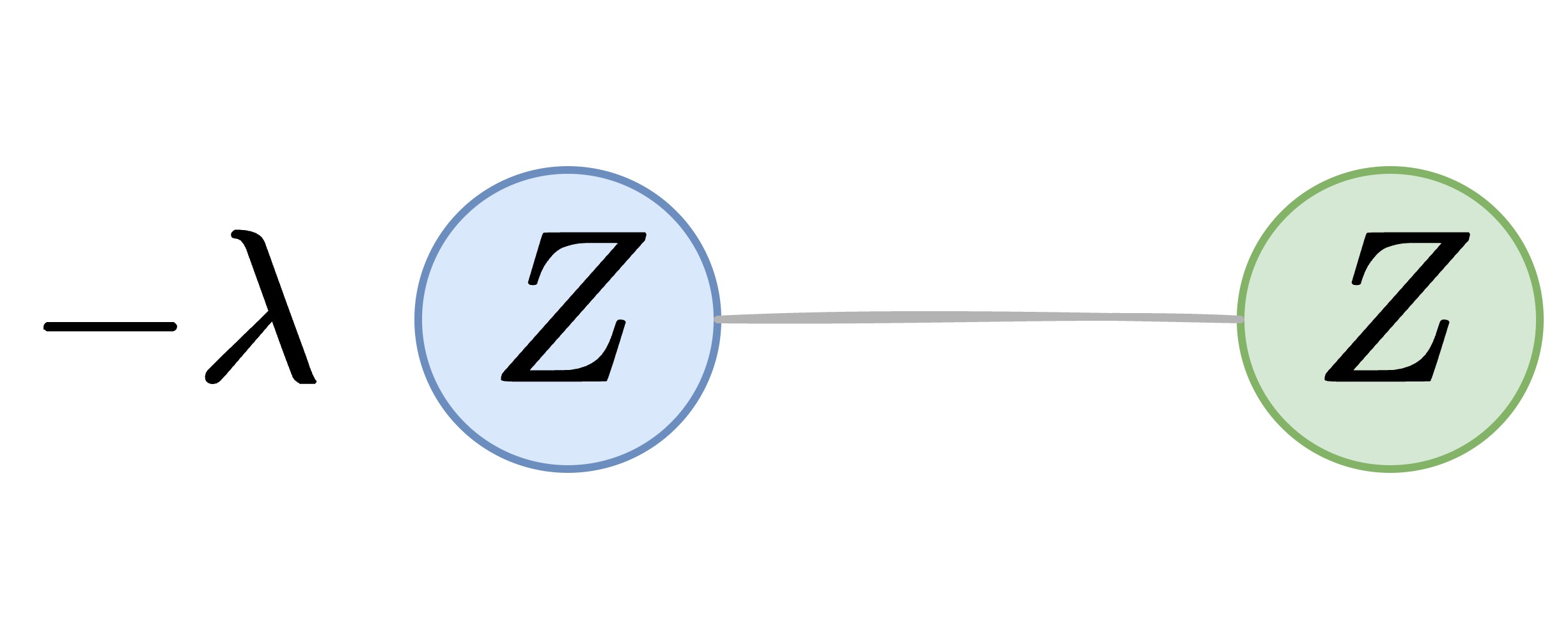}}},
\eeq
and tune coupling $\lambda\rightarrow\infty$. While this term explicitly break the $Z_2^e\times Z_2^o$ symmetry, the new ground state stabilizers can be given by the elements in the original stabilizer group that commute with the hopping term. Effectively, we can write the system on a square lattice, with ground state stabilizers,
\beq
    \vcenter{\hbox{\includegraphics[scale=.03,trim={0cm 0cm 0cm 0cm},clip]{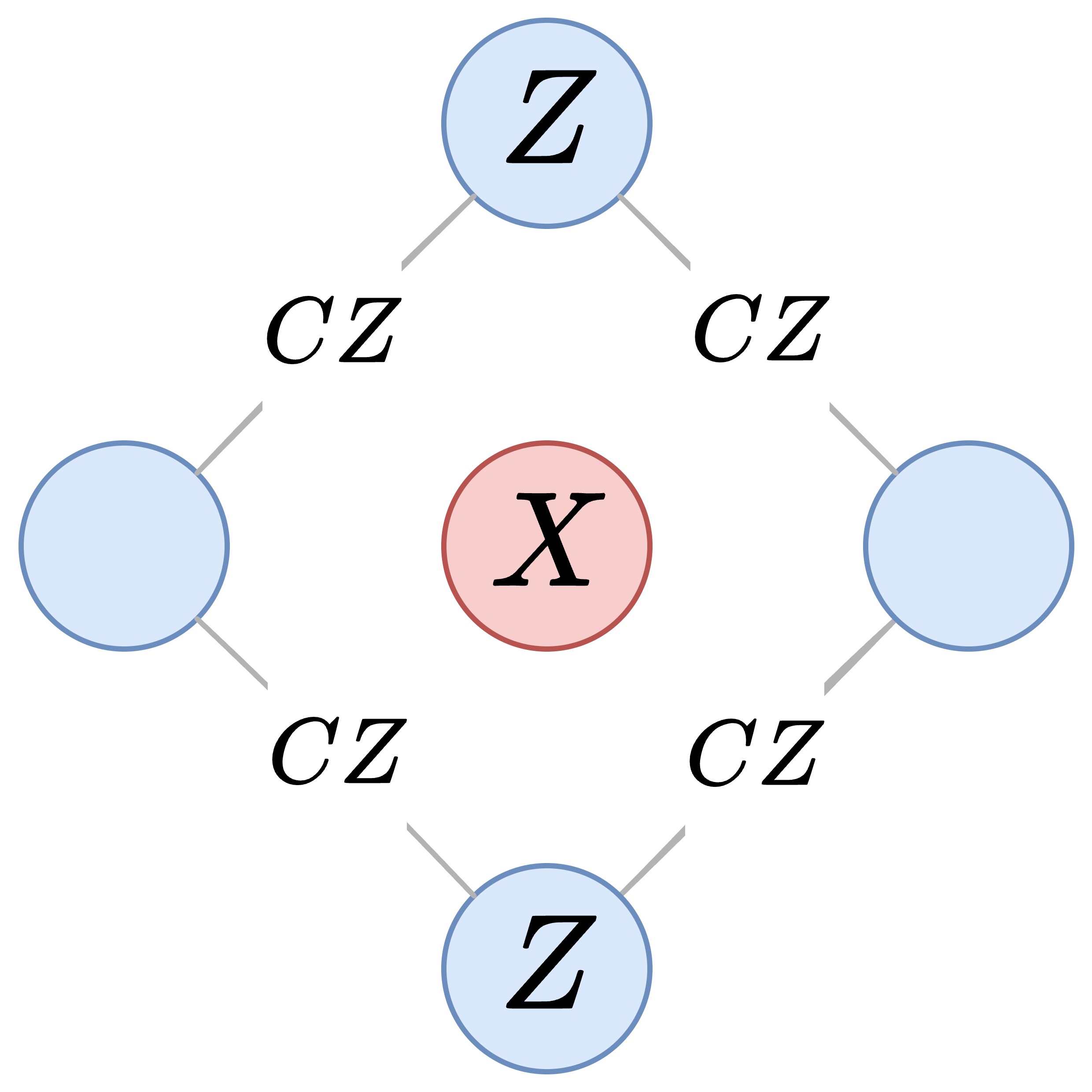}}},\ \vcenter{\hbox{\includegraphics[scale=.03,trim={0cm 0cm 0cm 0cm},clip]{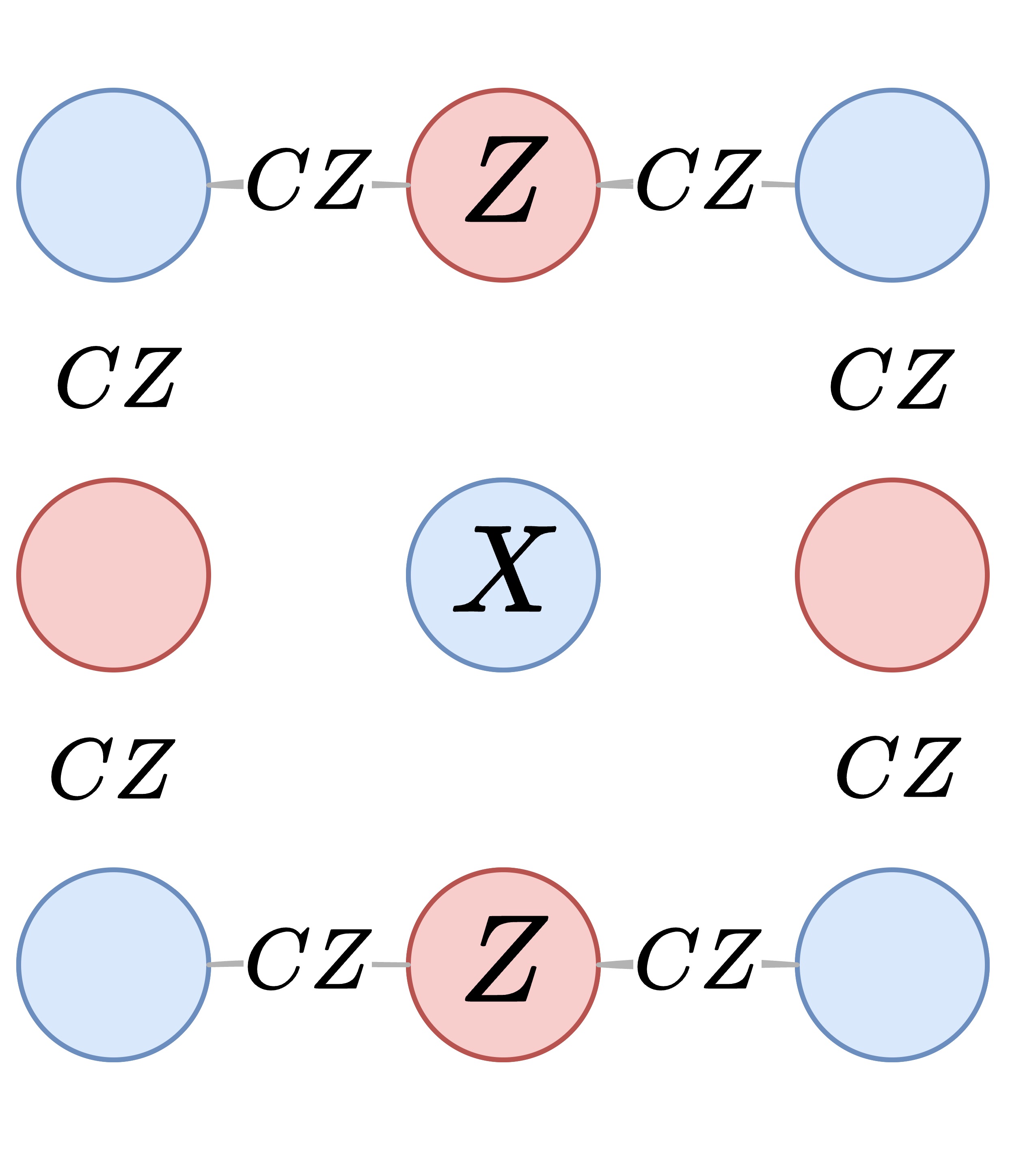}}}.
\eeq
As we discussed, this new ground state is a type II SPT state. Now suppose we want to write on a 3-colorable triangular lattice the type II+II+III SPT, which is a stack of three root states, we can obtain the SPT state by applying all three SPT entanglers on a product state. The stabilizers of this SPT are thus given in Fig.~\ref{fig:type-II+II+III}.

Now we consider the boundary of this SPT. We pick a boundary with $e$-sites and $o$-sites, on which there are $4L$ sites, and even sites are of color $e$, while odd sites are of color $o$. The boundary symmetry operators are those that commute with all bulk stabilizers. Therefore, they are given by,
\beq
\label{eq:SymType-II+II+III}
    U_e=\prod_j X_{2j},\ U_o=\prod_{j}X_{2j+1},\ V=\prod_{j}Z_j CZ_{j,j+1}CZ_{j,j+2}.
\eeq
These symmetry operators have a mixed type II+II+III anomaly as they can live on the boundary of such an SPT.

\section{$D_8$ anomaly from gauging Tambara-Yamagami categories}
\label{app:tachikawa}
In Ref.~\cite{meir2012module,bhardwaj2018finite}, it is established that, for a Tambara-Yamagami category TY($\mathcal{A,\chi,\tau}$), it can be obtained by gauging a subgroup of a possibly non-anomalous group, if and only if $\mathcal{A}$ has a Lagrangian subgroup $H$ for $\chi$, i.e. there is a subgroup $H \subset \mathcal{A}$ such that i) the restriction of $\chi$ on $H$ is trivial and ii) $\mathcal{A}/H\simeq \hat{H}$ via the pairing induced by $\chi$. 

We will first review the derivations in Sec.~5.6 of Ref.~\cite{bhardwaj2018finite}. For a TY category TY($\mathcal{A,\chi,\tau}$) satisfying the above conditions, the abelian group $\mathcal{A}$ is essentially a central extension of $\hat{H}$ by $H$ given by some 2-cocycle $\kappa$. The symmetric non-degenerate bicharacter $\chi$ defines a symmetric map, $\chi:\ \hat{H}\times \hat{H}\rightarrow U(1)$ satisfying the condition
\beq
    \chi(\rho+\rho',\sigma)=\chi(\rho,\sigma)\chi(\rho',\sigma)\sigma(\kappa(\rho,\rho')).
\eeq
Gauging the subgroup $H$, the symmetry operators are $X_{\rho,\sigma}$ and $Y_{\rho,\sigma}$, with $\rho,\sigma\in \hat{H}$. They form a group symmetry $(\hat{H}\times \hat{H})\rtimes Z_2$. The anomaly of this symmetry is in general given by a 3-cocycle $\alpha\in H^3((\hat{H}\times \hat{H})\rtimes Z_2,U(1))$,
\beq
    \alpha(X_{\rho,\sigma},X_{\rho',\sigma'},X_{\rho'',\sigma''})&=\rho''(\kappa(\sigma,\sigma')),\\
    \alpha(X_{\rho,\sigma},X_{\rho',\sigma'},Y_{\rho'',\sigma''})&=\rho''(\kappa(\sigma,\sigma')),\\
    \alpha(X_{\rho,\sigma},Y_{\rho',\sigma'},X_{\rho'',\sigma''})&=\chi(\sigma,\sigma''),\\
    \alpha(Y_{\rho,\sigma},X_{\rho',\sigma'},X_{\rho'',\sigma''})&=(\rho'\rho''\sigma^{-1})(\kappa(\sigma',\sigma'')),\\
    \alpha(X_{\rho,\sigma},Y_{\rho',\sigma'},Y_{\rho'',\sigma''})&=\sigma''(\kappa(\sigma,(\sigma')^{-1}\rho'')),\\
    \alpha(Y_{\rho,\sigma},X_{\rho',\sigma'},Y_{\rho'',\sigma''})&=\chi(\sigma',\sigma(\rho')^{-1}(\rho'')^{-1}),\\
    \alpha(Y_{\rho,\sigma},Y_{\rho',\sigma'},X_{\rho'',\sigma''})&=(\sigma'(\rho'')^{-1})(\kappa(\sigma(\rho')^{-1},\sigma'')),\\
    \alpha(Y_{\rho,\sigma},Y_{\rho',\sigma'},Y_{\rho'',\sigma''})&=\mathrm{sgn}(\tau)\chi(\sigma(\rho')^{-1},\sigma'(\rho'')^{-1}).
    \label{eq:3-cocycle_TY}
\eeq

When $\mathcal{A}=\Z_2\times \Z_2=\{1,a,b,ab\}$, there are four such TY categories corresponding to $\chi_{\pm}$ and $\tau=\pm \frac{1}{2}$ respectively, where the bicharacter is
\beq
    \chi_{\pm}(a,a)=\pm \chi_{\pm}(b,b)=\mp \chi_{\pm}(a,b)=\mp \chi_{\pm}(b,a)=1.
\eeq
Taking $H=\{1,a\}$, then after gauging $H$ from the TY categories, the symmetry becomes $\mathrm{Vec}_{D_8}^{\alpha}$. We embed the operators to $D_8=<a,x>$ under the conventions in Eq.~\eqref{eq:D_8} as
\beq
    X_{0,0} = 1, X_{0,1} =xa^2, X_{1,0}=x, X_{1,1}=a^2,\\
    Y_{0,0} = xa^3, Y_{0,1} =a, Y_{1,0}=a^3, Y_{1,1}=x a.
\eeq
It can be written straightforwardly from Eq.~\eqref{eq:3-cocycle_TY} that, for Rep($D_8$)$=$TY($\Z_2\times \Z_2,\chi_+,\frac{1}{2}$), the $D_8$ 3-cocycle after gauging is trivial
\beq
    \alpha_1(\tilde{g},\tilde{h},\tilde{k})\equiv 1.
\eeq 

For Rep($Q_8$)$=$TY($\Z_2\times \Z_2,\chi_+,-\frac{1}{2}$), the $D_8$ 3-cocycle after gauging is
\beq
    \alpha_2(\tilde{g},\tilde{h},\tilde{k})=(-1)^{ghk}.
\eeq
Indeed, this cocycle differs from the 3-cocycle $\omega$ in Eq.~\eqref{eq:Q_8_cocycle} by the coboundary of a 2-cochain
\beq
    \beta(\tilde{g},\tilde{h})=e^{\frac{\pi i}{2}[g]_2(H+h-[h]_2)},
\eeq
where $[g]_2:=g \text{ mod } 2$.

For Rep($H_8$)$=$TY($\Z_2\times \Z_2,\chi_-,\frac{1}{2}$), the $D_8$ 3-cocycle after gauging is
\beq
    \alpha_3(\tilde{g},\tilde{h},\tilde{k})=(-1)^{h\delta_{g,2}\delta_{k,2}+gk\delta_{h,2}\delta_{[g(-1)^{H+K}+h(-1)^K+k]_4,2}+ghk\delta_{[g(-1)^H+h]_4,2}\delta_{[h(-1)^K+k]_4,2}}.
\eeq
One can check that, by multiplying $\alpha_3$ with the coboundary of a 2-cochain
\beq
    \beta'(\tilde{g},\tilde{h})=e^{\frac{\pi i }{2}[g]_2(2H\delta_{h,2}+h-[h]_2)}(-1)^{\delta_{g,3}\delta_{h,2}},
\eeq
we can obtain the standard form of a $D_8$ 3-cocycle
\beq
    \omega'(\tilde{g},\tilde{h},\tilde{k})=e^{\frac{4\pi i }{16}g(-1)^{H+K}(h(-1)^K+k-[h(-1)^K+k]_4)}.
\eeq

\end{document}